\documentclass[12pt]{article}
\usepackage{amsmath}
\usepackage{amssymb}
\usepackage{amsthm}
\usepackage{graphicx}
\usepackage{subcaption}
\usepackage{enumerate}
\usepackage{natbib}
\usepackage{xcolor}
\usepackage{float}
\usepackage{multirow}
\usepackage{booktabs}
\usepackage{hyperref}
\hypersetup{colorlinks=true, linkcolor=blue, citecolor=blue, urlcolor=blue}

\newcommand{\blind}{0}

\newtheorem{theorem}{Theorem}
\newtheorem{proposition}[theorem]{Proposition}

\newtheorem{corollary}[theorem]{Corollary}
\theoremstyle{definition}
\newtheorem{definition}[theorem]{Definition}
\theoremstyle{remark}
\newtheorem{remark}[theorem]{Remark}

\newcommand{\E}{\mathbb{E}}
\newcommand{\R}{\mathbb{R}}
\newcommand{\Cov}{\mathrm{Cov}}
\newcommand{\Cohi}{\mathcal{C}}

\newcommand{\dgm}{\mathrm{dgm}}
\newcommand{\Lspl}{\Lambda}              
\newcommand{\sX}{\mathcal{X}}
\DeclareMathOperator*{\kmax}{kmax}       

\begin{document}

\def\spacingset#1{\renewcommand{\baselinestretch}%
{#1}\small\normalsize} \spacingset{1}


\if0\blind
{
  \title{\bf Spectral Topological Data Analysis of Brain Signals}
  \author{
    Anass El-Yaagoubi\thanks{
      Corresponding author: \texttt{anass.bourakna@kaust.edu.sa}.}\hspace{.2cm}\\
    Statistics Program, King Abdullah University of Science and Technology \\
    and \\
    Shuhao Jiao \\
    Department of Biostatistics, City University of Hong Kong \\
    and \\
    Moo K.\ Chung \\
    Department of Biostatistics \& Medical Informatics,
    University of Wisconsin--Madison \\
    and \\
    Hernando Ombao \\
    Statistics Program, King Abdullah University of Science and Technology
  }
  \maketitle
} \fi

\if1\blind
{
  \bigskip\bigskip\bigskip
  \begin{center}
    {\LARGE\bf Spectral Topological Data Analysis of Brain Signals}
  \end{center}
  \medskip
} \fi


\bigskip
\begin{abstract}
    Topological analyses of brain functional connectivity usually reduce each pair of channels to a single scalar dependence, typically the Pearson correlation, and so cannot resolve the frequency-specific synchronisation that organises electrophysiology. We propose a topological summary that keeps the frequency information. The spectral landscape indexes the persistence landscape of \citet{PL_FIRST} by Fourier frequency, building each filtration from a coherence-based distance, so that it is a function of both the filtration scale and the frequency. It is Lipschitz-stable in the coherence matrix and feeds a functional two-sample test over a chosen frequency band, whose limiting null distribution and consistency follow from standard functional-data arguments. In simulations the test recovers a topological difference in the band where it lives while holding its nominal level under the null. Applied to electroencephalography from $53$ control and $51$ ADHD children, a global test rejects equality of the two groups' cycle topology at the $95\%$ level ($p = 0.019$); a band-by-band follow-up localises the difference to the $\gamma$ and $\theta$ bands, although none survives family-wise correction at this sample size. The pattern is consistent with the established role of these bands in ADHD.
\end{abstract}

\noindent%

\spacingset{1.45}

\section{Introduction}
\label{sec:intro}

Functional connectivity networks built from neural recordings are now a standard tool for studying how the brain coordinates activity across regions \citep{HUMAN_CONNECTOME, FUNC_EFEC_CONN_FRISTON, STRUCT_FUNCT_BRAIN_NETWORKS, STRUCT_FUNCT_CONN_REVIEW, ELYAAGOUBI_RAT_LFP}. These networks are usually built in three steps: choose a pairwise dependence measure between channels, threshold to a binary graph, and compute graph-theoretic summaries such as modularity, centrality, or path length \citep{GRAPHS_BRAIN_DATA, BRAIN_NETWORK_SCIENCE, BRAIN_NETWORKS_SUMMARY}. This approach suffers from at least two practical issues. The threshold is somewhat arbitrary, and the resulting graph summaries are known to be sensitive to it \citep{PROBLEM_THRESHOLDING_SW_NETWORKS, OPTIMAL_THRESHOLDING, THRESHOLDING_PROBLEM_EEG}. The dependence measure is also usually the Pearson correlation, which reduces the signal in two channels to a single number and discards information about the frequency at which they co-vary; this is a substantial loss in electrophysiological data, where rhythm-specific synchronisation carries clinical meaning \citep{COHERENCE_EEG, SPECTRAL_DEPENDENCE}.

\begin{figure}[H]
\centering
\includegraphics[width=0.8\linewidth]{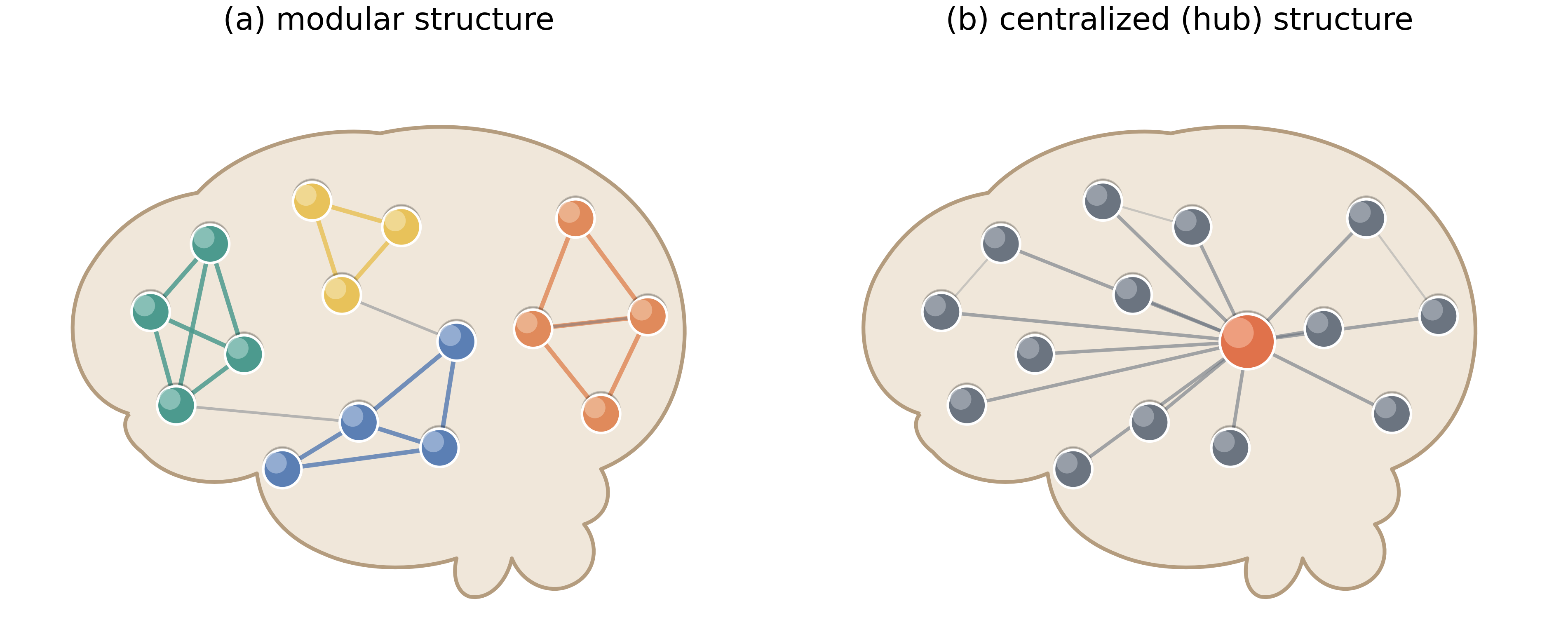}
\caption{Two toy examples of brain connectivity at a single threshold: a modular network (left) and a hub network (right). Graph-theoretic summaries such as modularity and centrality can distinguish them here, but their values depend on the threshold and the same networks can look very different at other thresholds.}
\label{fig:centrality_vs_modularity}
\end{figure}

\begin{figure}[H]
\centering
\includegraphics[width=0.9\linewidth]{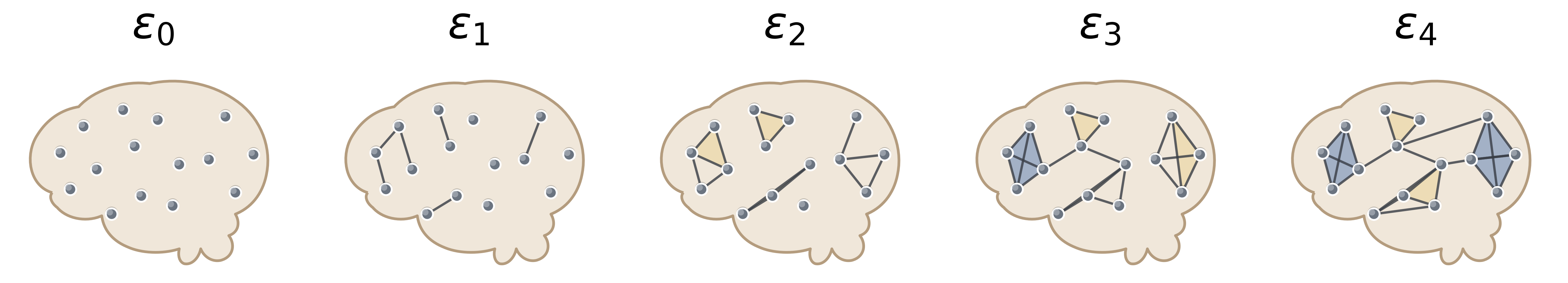}
\caption{A network filtration: as the scale grows (left to right), simplices
are added and topological features appear and disappear.}
\label{fig:brain_filtration}
\end{figure}

Graph-theoretic summaries such as modularity and node centrality have been successful in characterising brain networks and can distinguish patterns like those in Figure~\ref{fig:centrality_vs_modularity}, but they require committing to a single threshold. Persistent homology instead examines the network across all thresholds simultaneously by building a filtration (Figure~\ref{fig:brain_filtration}); we provide more detail in Section~\ref{subsec:coherence-distance}. One of the first persistent-homology analyses of brain networks is due to \citet{TDA_WEIGHTED_NETWORKS}, and the approach has since uncovered structural and disease-related features that single-threshold graph analyses miss \citep{CAPUTI_TDA_BRAIN_CONNECTIVITY, TDA_NEUROSCIENCE, GRAPH_THEORY_NEUROLOGICAL_DISORDERS, TDA_MULTIVARIATE_TS_ANASS}.

\begin{figure}[H]
\centering
\includegraphics[width=0.9\linewidth]{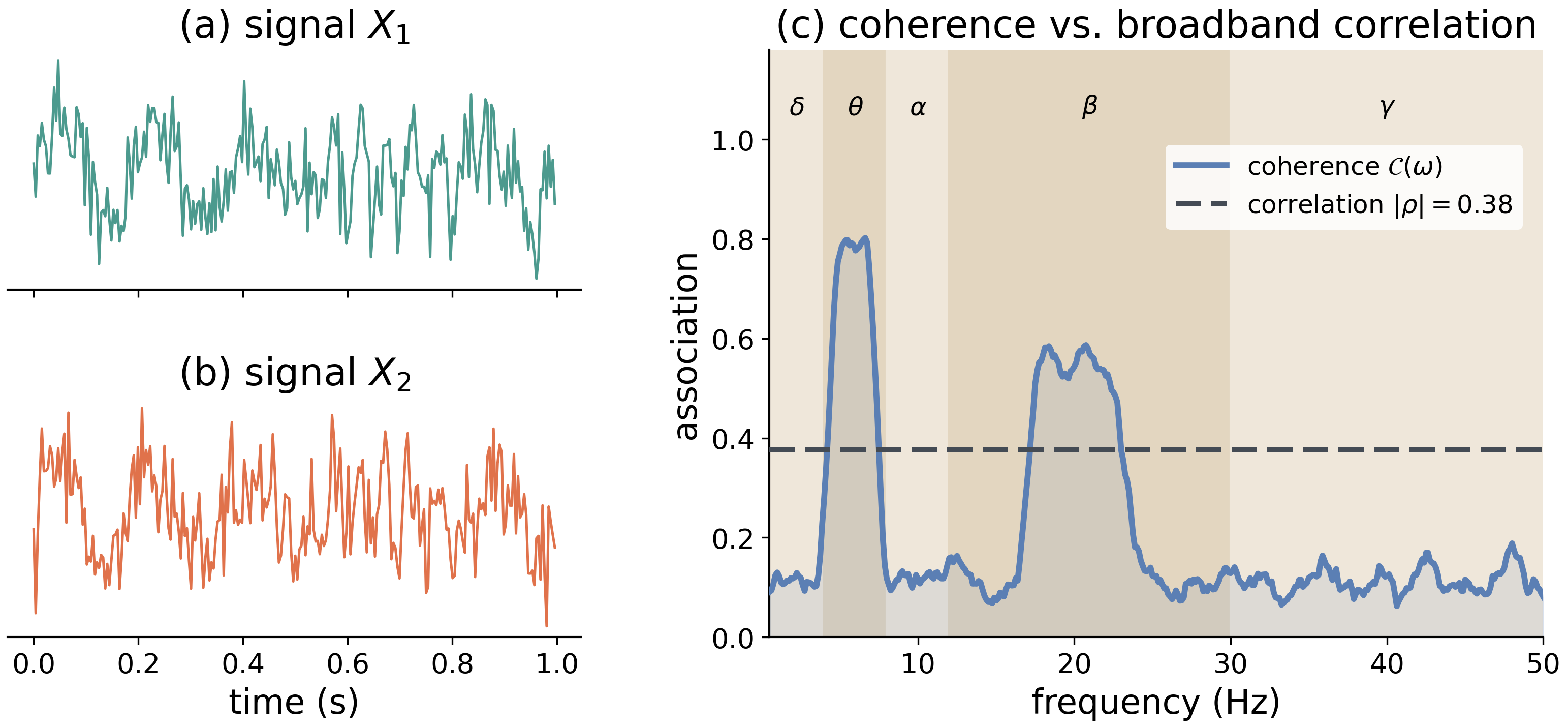}
\caption{Coherence captures frequency-specific dependence a single
correlation cannot. (a,b)~Two synthetic signals coupled in $\theta$ and
$\beta$ bands. (c)~Their coherence $\mathcal{C}(\omega)$ versus broadband $|\rho|$
on a common axis.}
\label{fig:coherence_signal_decomposition}
\end{figure}

These topological pipelines, however, inherit the frequency problem of the underlying dependence measure: built on correlation-based distances, they cannot distinguish patterns that live at different oscillatory frequencies. For electroencephalography (EEG), where the $\delta$, $\theta$, $\alpha$, $\beta$, and $\gamma$ bands have well-established cognitive and clinical roles \citep{BRAIN_NETWORKS_BOOK, GRAPH_THEORY_BRAIN_NETWORKS}, this is a real loss of information. Figure~\ref{fig:coherence_signal_decomposition} illustrates the point: coherence resolves the frequency-specific dependence between two signals that a single broadband correlation collapses to one number. Coherence in general has a long history in EEG analysis \citep{COHERENCE_EEG, COHERENCE_HIST, SPECTRAL_DEPENDENCE}. Beyond stationary signals, topological methods have also been applied to non-stationary signals through sliding-window embeddings \citep{SLIDING_WINDOW_TDA, TDA_TSA, DYNAMIC_TDA_ANASS}.

The present paper differs in the role assigned to frequency. We replace the scalar dependence measure with the coherence at frequency $\omega$, which produces a family of distance matrices indexed by $\omega$, for which a frequency-indexed filtration is a natural fit. The associated topological summary, which we call the \emph{spectral landscape}, is therefore a function of both the filtration scale $s$ and the frequency $\omega$. This construction is also technically convenient: it yields a family of one-parameter filtrations rather than a multi-parameter persistence module, so the standard machinery of barcodes, landscapes, and stability applies frequency by frequency.

The spectral landscape $\Lambda_k(s,\omega)$ extends the persistence landscape of \citet{PL_FIRST} by indexing it with frequency, giving a topological summary that lives in a separable Banach space and inherits the usual statistical machinery: a well-defined population mean, a strong law of large numbers, and a central limit theorem. The summary is Lipschitz-stable in the underlying coherence matrix, so small errors in the estimated coherence translate into small errors in the landscape. We use it as input to a functional two-sample test for population-mean differences over a chosen frequency band, with the asymptotic theory inherited from \citet{PL_FIRST} and \citet{horvath2012inference}, and we address the multiple comparisons across frequency bands explicitly.

Section~\ref{sec:meth} develops the methodology: the coherence-based distance and frequency-indexed filtration, the spectral landscape, its asymptotic properties, the stability bound, and the two-sample test. Section~\ref{sec:sim} reports the simulation studies, where the spectral landscape recovers the band-localised topology built into four data-generating settings and the test both discriminates those settings and holds its nominal level under the null. Section~\ref{sec:app} applies the method to the ADHD EEG data, and Section~\ref{sec:conc} concludes. Background material on persistence homology and the proofs are collected in the Supplemental Materials.

\section{Methods}
\label{sec:meth}

\subsection{Coherence and frequency-specific distance}
\label{subsec:coherence-distance}

Let $X(t) = (X_1(t), \ldots, X_P(t))^\top$ denote a zero-mean, second-order stationary $P$-variate time series observed at times $t = 1, \ldots, T$. Here $P$ is the number of recording channels (brain locations) and $T$ is the number of time points. Under stationarity, $X(t)$ admits the Cram\'er spectral representation
\begin{equation}
    X(t) = \int_{-1/2}^{1/2} \exp(i 2\pi \omega t)\, dZ(\omega),
    \label{eq:cramer}
\end{equation}
where $dZ(\omega) = (dZ_1(\omega), \ldots, dZ_P(\omega))^\top$ is a $P$-variate orthogonal-increment process satisfying $\E[dZ(\omega)] = 0$ and $\Cov(dZ(\omega), dZ(\omega')) = 0$ for $\omega \neq \omega'$. The spectral density matrix $S(\omega) \in \mathbb{C}^{P \times P}$ is defined by
\begin{equation}
    S(\omega)\, d\omega \;=\; \E\!\left[ dZ(\omega)\, dZ(\omega)^{*} \right],
    \label{eq:spectral-matrix}
\end{equation}
where $*$ denotes conjugate transpose. For $p, q \in \{1, \ldots, P\}$, the entry $S_{pq}(\omega)$ is the cross-spectrum between channels $p$ and $q$, and the diagonal entries $S_{pp}(\omega)$ are the auto-spectra. The coherence between channels $p$ and $q$ at frequency $\omega$ is
\begin{equation}
    \Cohi_{pq}(\omega) \;=\; \frac{|S_{pq}(\omega)|}{\sqrt{S_{pp}(\omega)\, S_{qq}(\omega)}} \;\in\; [0,1].
    \label{eq:coherence}
\end{equation}
Coherence quantifies the strength of linear association between the $\omega$-oscillatory components of $X_p$ and $X_q$ and is interpretable as the absolute correlation between the band-pass--filtered components at frequency $\omega$ \citep{SPECTRAL_DEPENDENCE}. It is widely used in EEG analysis precisely because brain signals exhibit frequency-band-specific synchronization that scalar correlation cannot resolve \citep{COHERENCE_EEG, COHERENCE_HIST}.

Given an estimator $\widehat{\Cohi}_{pq}(\omega)$ of $\Cohi_{pq}(\omega)$ (a smoothed-periodogram estimator, with bandwidth specified in Section~\ref{subsec:estimation}), we define the frequency-specific distance
\begin{equation}
    d_X(X_p, X_q, \omega) \;=\; 1 - \widehat{\Cohi}_{pq}(\omega) \;\in\; [0,1].
    \label{eq:freq-distance}
\end{equation}
The construction is generic: any frequency-specific dependence measure taking values in $[0,1]$ can be substituted for $\widehat{\Cohi}_{pq}(\omega)$, and any strictly decreasing link $\mathcal{G}: [0,1] \to [0,1]$ can replace $1 - (\cdot)$.
For each frequency $\omega$, the pair $(\{1, \ldots, P\}, d_X(\cdot, \cdot, \omega))$ is a finite pseudo-metric space\footnote{The triangle inequality follows from $\widehat{\Cohi}_{pq}(\omega)$ being the cosine of an angle in the Cram\'er Hilbert-space representation of the channels at frequency $\omega$; it becomes a strict metric once channels with $\widehat{\Cohi}_{pq}(\omega) = 1$ are identified.}, on which we construct the Vietoris-Rips filtration
\begin{equation}
    \sX_{\epsilon_1}(\omega) \subset \sX_{\epsilon_2}(\omega) \subset \cdots \subset \sX_{\epsilon_n}(\omega),
    \label{eq:filtration}
\end{equation}
where $0 = \epsilon_1 < \epsilon_2 < \cdots < \epsilon_n = 1$ is an increasing grid of distance thresholds and $\sX_\epsilon(\omega)$ is the simplicial complex whose $k$-simplices are subsets of $k+1$ channels pairwise within distance $\epsilon$ at frequency $\omega$. We refer the reader to \citet{MUNKRES_ALG_TOP} and \citet{HAUSMANN_RIPS_FILTRATION} for the formal construction of simplicial complexes and Vietoris-Rips complexes over metric spaces, and to \citet{BARCODES} for an applied introduction; self-contained background tailored to this paper is in Supplement~A. The construction in~\eqref{eq:filtration} yields one filtration per frequency; the topology of brain dependence is now a frequency-indexed object, not a scalar.

For numerical implementation we use the uniform Fourier grid $\omega_\ell = \ell f_s / T$ for $\ell = 1, \ldots, \lfloor T/2 \rfloor$, where $f_s$ is the sampling frequency in Hertz. With $T$ time points and $P$ channels, the cost of building the family of distance matrices is $\mathcal{O}(P T \log T + \lfloor T/2 \rfloor \cdot P^{2})$ (one FFT per channel, then per-frequency cross-spectra and coherences). Persistent homology is then computed independently at each frequency, which is polynomial in $P$ for the dimensions $k = 0, 1$ considered here and computationally light at the channel counts typical in EEG.

\subsection{Spectral landscape}
\label{subsec:spectral-landscape}

For each frequency $\omega$, persistent homology of the filtration~\eqref{eq:filtration} in dimension $k$ produces a persistence diagram $\dgm_k(\omega) = \{(b_j, d_j)\}_{j=1}^{m(\omega)}$, where each pair $(b_j, d_j) \in [0,1]^2$ with $b_j < d_j$ records the filtration scale at which a $k$-dimensional homological feature is born and dies (Supplement~A). To enable formal inference, we summarize this diagram via persistence landscapes \citep{PL_FIRST}: for each pair $(b, d) \in \dgm_k(\omega)$, define the tent function
\begin{equation}
\lambda^{(b,d)}(s) =
\begin{cases}
s - b, & s \in [b, (b+d)/2), \\
d - s, & s \in [(b+d)/2, d), \\
0, & \text{otherwise.}
\end{cases}
\label{eq:tent}
\end{equation}
The persistence landscape at frequency $\omega$ is then the sequence of pointwise maxima of these tent functions.

\begin{definition}[Spectral landscape]
\label{def:spectral-landscape}
For homology dimension $k \ge 0$ and level $\ell \ge 1$, the \emph{$\ell$-th spectral landscape} is the function $\Lspl_{k,\ell} : [0,1] \times [0, 1/2] \to \R$ defined by
\begin{equation}
\Lspl_{k,\ell}(s, \omega)
\;=\; \kmax_{(b,d) \in \dgm_k(\omega)}\; \lambda^{(b,d)}(s),
\qquad s \in [0,1],\ \omega \in [0, 1/2],
\label{eq:spectral-landscape}
\end{equation}
where $\kmax_\ell$ denotes the $\ell$-th largest value (ties broken arbitrarily).
\end{definition}

\begin{figure}[H]
\centering
\includegraphics[width=.55\linewidth]{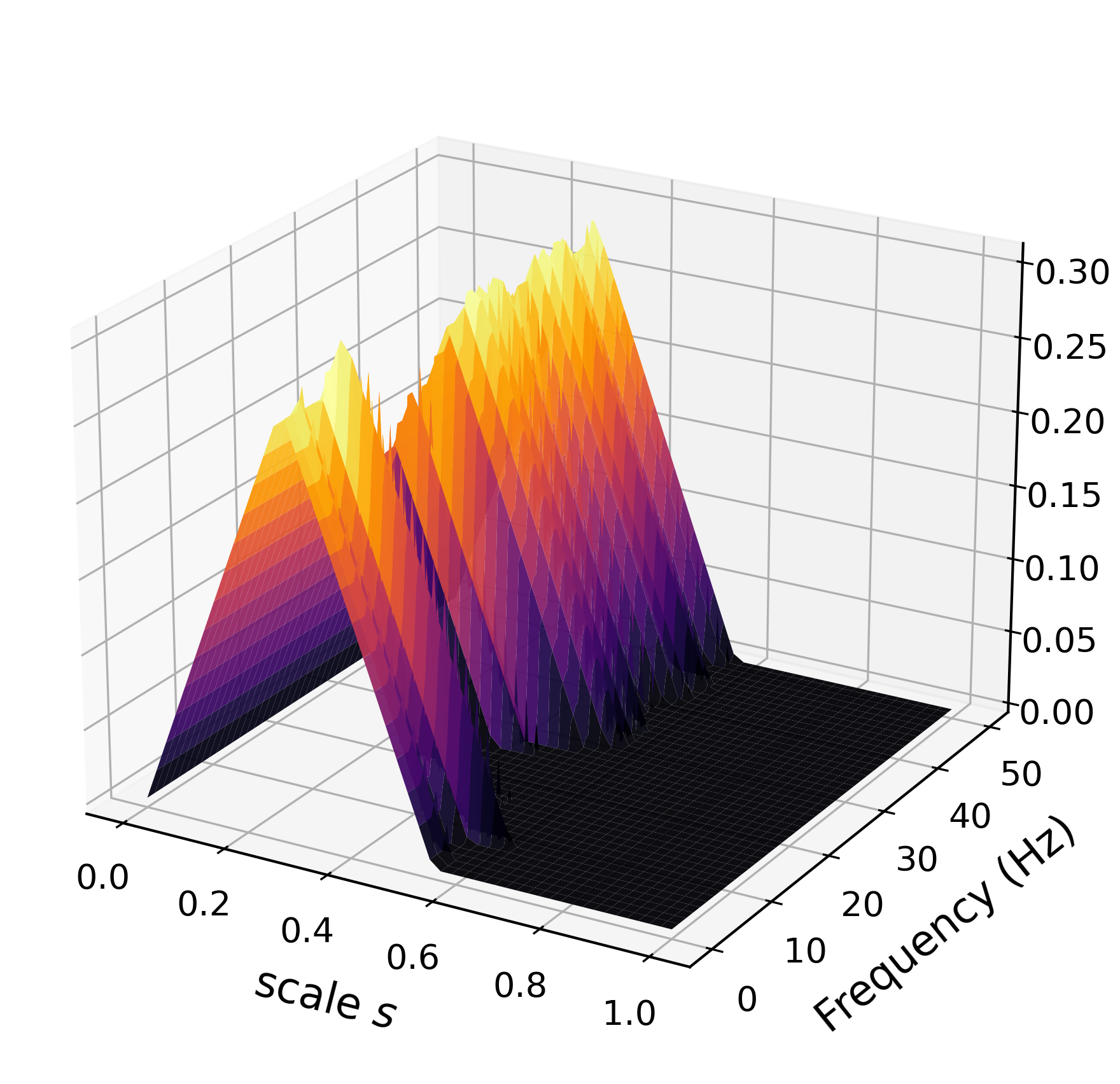}
\caption{Example of a spectral landscape $\Lspl_0(s, \omega)$ corresponding to the $0$-dimensional homology group. The horizontal axes are the filtration scale $s$ and frequency $\omega$ (in Hz); the vertical axis is the landscape height.}
\label{fig:spectral_landscape_example}
\end{figure}

\noindent For each fixed $\omega$, $\Lspl_{k,\ell}(\cdot, \omega)$ is the $\ell$-th persistence landscape of $\dgm_k(\omega)$ in the sense of \citet{PL_FIRST}; an efficient algorithmic implementation is described in \citet{PL_COMPUTE}. The frequency argument $\omega$ promotes a single landscape to a 2D surface; we call this surface the spectral landscape because it indexes the topological summary by the spectral variable. The level $\ell$ controls which topological feature one tracks: $\ell = 1$ corresponds to the most persistent feature at each $s$, $\ell = 2$ to the second-most persistent, and so on. Unless stated otherwise we work with $\ell = 1$ and write $\Lspl_k(s, \omega) := \Lspl_{k,1}(s, \omega)$. Figure~\ref{fig:spectral_landscape_example} shows an example for $k = 0$. When the spectral landscape is computed from data of subject $i$, we write $\Lspl_k^{(i)}(s,\omega)$. The superscript $i$ refers to subject; the subscript $k$ to homology dimension.

\medskip
\noindent The following properties follow directly from Definition~\ref{def:spectral-landscape} and the fact that birth and death values lie in $[0,1]$:
\begin{enumerate}
\item[\textup{(P1)}] $0 \le \Lspl_{k,\ell}(s,\omega) \le 1/2$ for all $s, \omega$;
\item[\textup{(P2)}] $\Lspl_{k,\ell}(s,\omega) \ge \Lspl_{k,\ell+1}(s,\omega)$ for all $\ell \ge 1$;
\item[\textup{(P3)}] $\Lspl_{k,\ell}(\cdot, \omega)$ is $1$-Lipschitz in $s$ for every fixed $\omega$.
\end{enumerate}

\subsection{Asymptotic properties}
\label{subsec:asymptotics}

In a study with $n$ independent subjects, each subject $i$ yields an estimated coherence matrix $\widehat{\Cohi}^{(i)}(\omega)$ and, via Definition~\ref{def:spectral-landscape}, a spectral landscape $\Lspl_k^{(i)}(s, \omega)$. We view $\Lspl_k^{(i)}$ as a Borel random element of the separable Banach space $L^p([0,1] \times [0,1/2])$, $1 \le p < \infty$, with norm $\|\cdot\|_p$. Write $\overline{\Lspl}_k^n = n^{-1} \sum_{i=1}^n \Lspl_k^{(i)}$ for the sample mean and $\mu_k = \E[\Lspl_k^{(i)}]$ for the population mean.

The persistence-landscape framework of \citet{PL_FIRST} establishes that the strong law of large numbers and the central limit theorem for Banach-valued random variables \citep{LLN_CLT, PROBABILITY_BANACH_SPACES} apply directly to persistence landscapes. The same arguments apply unchanged to spectral landscapes, since for each fixed $\omega$ they are persistence landscapes in the sense of Bubenik and the additional frequency argument enters only as an index in a separable Banach space. We record the consequences for completeness; the proof is a direct application of \citet{PL_FIRST} and is given in Supplement~B.

\begin{proposition}[Limit theorems for spectral landscapes]
\label{prop:limits}
Let $\Lspl_k^{(1)}, \Lspl_k^{(2)}, \ldots$ be i.i.d.\ copies of a Borel random element $\Lspl_k$ of $L^p([0,1] \times [0,1/2])$, $1 \le p < \infty$.
\begin{enumerate}
\item[\textup{(i)}] If $\E\|\Lspl_k\|_p < \infty$, then $\overline{\Lspl}_k^n \to \mu_k$ almost surely as $n \to \infty$.
\item[\textup{(ii)}] If $\E\|\Lspl_k\|_p^2 < \infty$, then $\sqrt{n}\,(\overline{\Lspl}_k^n - \mu_k)$ converges weakly to a centered Gaussian element of $L^p([0,1] \times [0,1/2])$ with the same covariance as $\Lspl_k$.
\end{enumerate}
\end{proposition}

\noindent The boundedness of $\Lspl_k$ ($0 \le \Lspl_k \le 1/2$ by (P1)) on a compact domain guarantees the moment conditions for any $p < \infty$, so both conclusions hold without additional assumptions in our setting.

\begin{remark}[Estimation error in coherence]
\label{rem:est-error}
Proposition~\ref{prop:limits} treats $\Lspl_k^{(i)}$ as i.i.d.\ across subjects, which absorbs into the population-level variability of $\Lspl_k$ both (a) genuine between-subject variation in the underlying coherence structure and (b) within-subject estimation error in $\widehat{\Cohi}^{(i)}$. This is the standard two-stage modeling choice in functional data analysis \citep{horvath2012inference}. The stability bound in Section~\ref{subsec:stability} quantifies how within-subject estimation error in $\widehat{\Cohi}^{(i)}$ propagates to $\Lspl_k^{(i)}$, making this absorption explicit and bounded.
\end{remark}

\subsection{Stability of spectral landscapes}
\label{subsec:stability}

A practical concern with any topological summary built from estimated quantities is whether small errors in the estimate produce only small errors in the summary. For spectral landscapes, this question reduces to two well-known stability results that we chain together.

Fix a frequency $\omega$ and consider two coherence matrices $\Cohi(\omega), \Cohi'(\omega) \in [0,1]^{P \times P}$, with corresponding distance matrices $D(\omega) = 1 - \Cohi(\omega)$ and $D'(\omega) = 1 - \Cohi'(\omega)$ (taken entry-wise, with the diagonal set to zero). Let $\dgm_k(\omega)$ and $\dgm_k'(\omega)$ be the persistence diagrams in dimension $k$ of the Vietoris-Rips filtrations built from $D(\omega)$ and $D'(\omega)$, respectively, and let $\Lspl_{k,\ell}(\cdot, \omega)$ and $\Lspl_{k,\ell}'(\cdot, \omega)$ be the corresponding $\ell$-th persistence landscapes.

\begin{theorem}[Pointwise stability in frequency]
\label{thm:stability-pointwise}
For every $\omega \in [0, 1/2]$, every homology dimension $k \ge 0$, and every level $\ell \ge 1$,
\begin{equation}
\left\| \Lspl_{k,\ell}(\cdot, \omega) - \Lspl_{k,\ell}'(\cdot, \omega) \right\|_\infty
\;\le\;
\max_{1 \le p,q \le P} \left| \Cohi_{pq}(\omega) - \Cohi'_{pq}(\omega) \right|.
\label{eq:stab-pointwise}
\end{equation}
\end{theorem}

\begin{proof}[Sketch]
By the bottleneck stability theorem for persistence diagrams of filtered simplicial complexes \citep{PD_STABILITY}, the bottleneck distance between $\dgm_k(\omega)$ and $\dgm_k'(\omega)$ is bounded by the $\ell_\infty$-distance between the distance matrices $D(\omega)$ and $D'(\omega)$. Since $D - D' = -(\Cohi - \Cohi')$ entry-wise, this $\ell_\infty$-distance equals $\max_{p,q} |\Cohi_{pq}(\omega) - \Cohi'_{pq}(\omega)|$. The persistence landscape is $1$-Lipschitz with respect to the bottleneck distance in the sense that $\|\Lspl_{k,\ell} - \Lspl_{k,\ell}'\|_\infty \le W_\infty(\dgm_k, \dgm_k')$ \citep[Theorem~12]{PL_FIRST}, where $W_\infty$ is bottleneck distance. Chaining the two inequalities gives~\eqref{eq:stab-pointwise}. Full details are in Supplement~B.
\end{proof}

The bound is uniform in $\omega$: integrating over any frequency band $\Omega_\ell \subset [0, 1/2]$ yields the following $L^\infty$-in-$s$, $L^2$-in-$\omega$ control.

\begin{corollary}[Spectral-band stability]
\label{cor:stability-band}
For any frequency band $\Omega_\ell \subset [0, 1/2]$,
\begin{equation}
\int_{\Omega_\ell} \left\| \Lspl_{k,\ell}(\cdot, \omega) - \Lspl_{k,\ell}'(\cdot, \omega) \right\|_\infty^2 d\omega
\;\le\;
\int_{\Omega_\ell} \max_{p,q} \left| \Cohi_{pq}(\omega) - \Cohi'_{pq}(\omega) \right|^2 d\omega.
\label{eq:stab-band}
\end{equation}
\end{corollary}

\begin{remark}[Consequence for inference]
Theorem~\ref{thm:stability-pointwise} and Corollary~\ref{cor:stability-band} make precise the sense in which the spectral landscape inherits the consistency of $\widehat{\Cohi}$: if $\widehat{\Cohi}^{(i)}(\omega) \to \Cohi^{(i)}(\omega)$ uniformly in $\omega$ as $T \to \infty$ (which holds for the smoothed periodogram under standard mixing and bandwidth conditions; see, e.g., \citet{SPECTRAL_DEPENDENCE} and references therein), then the estimated spectral landscape converges to its population counterpart at the same uniform rate. This justifies the use of $\Lspl_k^{(i)}$ as an input to the asymptotic theory of Proposition~\ref{prop:limits} with the standard FDA caveat (Remark~\ref{rem:est-error}).
\end{remark}

\subsection{A functional two-sample test for frequency-specific topological differences}
\label{subsec:test}

Spectral landscapes are 2D functional summaries, and comparison of two populations of subjects (e.g., control vs.\ disease) is naturally cast as a functional two-sample test \citep{horvath2012inference}. Let $\Pi_1$ and $\Pi_2$ denote two populations, and fix a homology dimension $k$; for brevity we drop $k$ from the notation throughout this subsection. Assume the model
\begin{equation*}
\Lspl^{(i)}(s, \omega) =
\begin{cases}
\mu_1(s, \omega) + \mathcal{E}^{(i)}(s, \omega), & i \in \Pi_1, \\
\mu_2(s, \omega) + \mathcal{E}^{(i)}(s, \omega), & i \in \Pi_2,
\end{cases}
\end{equation*}
where the error functions $\mathcal{E}^{(i)}$ are i.i.d.\ within each population, mean zero, and have finite second moments. We write
\begin{align}
\Gamma_g(s_1, \omega_1, s_2, \omega_2)
&= \E\!\left[ \mathcal{E}^{(i)}(s_1, \omega_1)\, \mathcal{E}^{(i)}(s_2, \omega_2) \right],
\qquad i \in \Pi_g,\ g \in \{1, 2\},
\label{eq:Gamma_g}
\end{align}
for the population-$g$ covariance kernel.

Given a frequency band of interest $\Omega_\ell \subset [0, 1/2]$, we test
\begin{equation}
H_0 : \mu_1(s, \omega) = \mu_2(s, \omega) \quad \text{for all } s \in [0,1],\ \omega \in \Omega_\ell,
\qquad \text{vs.\ } H_1: \text{not } H_0.
\label{eq:H0}
\end{equation}
Let $N_1$ and $N_2$ be the sample sizes from $\Pi_1$ and $\Pi_2$, with sample means $\widehat{\mu}_1, \widehat{\mu}_2$, and suppose $N_1 / (N_1 + N_2) \to c \in (0, 1)$ as $N_1, N_2 \to \infty$. The test statistic is the integrated squared difference of sample means over $[0,1] \times \Omega_\ell$:
\begin{equation}
T^{\Omega_\ell}_{N_1, N_2}
\;=\; \frac{N_1 N_2}{N_1 + N_2} \int_{[0,1] \times \Omega_\ell} \left(\widehat{\mu}_1 - \widehat{\mu}_2\right)^2 (s, \omega)\, ds\, d\omega.
\label{eq:teststat}
\end{equation}

\begin{theorem}[Asymptotic null distribution]
\label{thm:test}
Under $H_0$ and the moment condition $\E\|\mathcal{E}^{(i)}\|_2^2 < \infty$,
\begin{equation}
T^{\Omega_\ell}_{N_1, N_2}
\;\xrightarrow{d}\;
\int_{[0,1] \times \Omega_\ell} G^2(s, \omega)\, ds\, d\omega
\;=\; \sum_{d=1}^{\infty} \lambda_d\, N_d^2,
\label{eq:null-dist}
\end{equation}
where $G$ is a centered Gaussian process on $[0,1] \times \Omega_\ell$ with covariance
\begin{equation}
\Gamma(s_1, \omega_1, s_2, \omega_2) = (1 - c)\, \Gamma_1(s_1, \omega_1, s_2, \omega_2) + c\, \Gamma_2(s_1, \omega_1, s_2, \omega_2),
\label{eq:pooled-cov}
\end{equation}
$\{\lambda_d\}_{d \ge 1}$ are the eigenvalues of the integral operator with kernel $\Gamma$, and $\{N_d\}_{d \ge 1}$ are i.i.d.\ standard normal.
\end{theorem}

\begin{proof}[Sketch]
By Proposition~\ref{prop:limits}, $\sqrt{N_g}(\widehat{\mu}_g - \mu_g) \xrightarrow{d} G_g$ for $g = 1, 2$, with $G_1, G_2$ independent centered Gaussian elements of covariance $\Gamma_1, \Gamma_2$. Writing
\begin{equation*}
\sqrt{\tfrac{N_1 N_2}{N_1 + N_2}}(\widehat{\mu}_1 - \widehat{\mu}_2)
= \sqrt{\tfrac{N_2}{N_1 + N_2}}\sqrt{N_1}(\widehat{\mu}_1 - \mu_1)
- \sqrt{\tfrac{N_1}{N_1 + N_2}}\sqrt{N_2}(\widehat{\mu}_2 - \mu_2)
\end{equation*}
and applying the continuous mapping theorem with $N_1/(N_1+N_2) \to c$ yields convergence to $\sqrt{1-c}\,G_1 - \sqrt{c}\,G_2$, which is centered Gaussian with covariance $\Gamma$ in~\eqref{eq:pooled-cov}. The Karhunen--Lo\`eve expansion of $G$ gives the chi-square mixture representation. Full details are in Supplement~B and follow \citet[Ch.~5]{horvath2012inference}.
\end{proof}

In practice $c$, $\Gamma_1$, $\Gamma_2$ are unknown and are estimated by $\widehat{c} = N_1 / (N_1 + N_2)$ and the empirical covariances
\begin{equation*}
\widehat{\Gamma}_g(s_1, \omega_1, s_2, \omega_2)
= \tfrac{1}{N_g} \sum_{i \in \Pi_g} \left(\Lspl^{(i)} - \widehat{\mu}_g\right)(s_1, \omega_1)\, \left(\Lspl^{(i)} - \widehat{\mu}_g\right)(s_2, \omega_2),
\quad g \in \{1, 2\},
\end{equation*}
combined as $\widehat{\Gamma} = (1 - \widehat{c})\widehat{\Gamma}_1 + \widehat{c}\,\widehat{\Gamma}_2$. The eigenvalues $\{\widehat{\lambda}_d\}$ are obtained by discretizing $\widehat{\Gamma}$ on the $(s, \omega)$ grid and computing the spectrum of the resulting matrix. Since the rank of $\widehat{\Gamma}$ is at most $N_1 + N_2 - 2$, the chi-square mixture in~\eqref{eq:null-dist} can be truncated at $D = N_1 + N_2 - 2$ without loss. Critical values and $p$-values are then obtained by Monte Carlo simulation from $\sum_{d=1}^{D} \widehat{\lambda}_d N_d^2$.

When the test in~\eqref{eq:H0} is applied to several disjoint frequency bands $\Omega_{\ell_1}, \ldots, \Omega_{\ell_B}$ and/or several homology dimensions $k$, the resulting $p$-values must be corrected for multiplicity. In the application of Section~\ref{sec:app} we report raw, Bonferroni-corrected, and Benjamini--Hochberg-adjusted $p$-values across the five canonical EEG bands ($\delta, \theta, \alpha, \beta, \gamma$) and the two homology dimensions ($k = 0, 1$), giving a family size of $B = 10$; the simulation study of Section~\ref{sec:sim} applies the same corrections to its own family of pairwise tests.

\section{Simulation Studies}
\label{sec:sim}

The simulations have two purposes. The first is to confirm that the spectral landscape encodes what it is designed to encode, namely topological structure that is localised at a particular frequency. The second is to check that the two-sample test of Section~\ref{subsec:test} holds its nominal level when the two groups come from the same process.

We build the data-generating processes around four examples. Two settings share a connectivity topology but place it in different frequency bands, and two settings share a band but carry different topologies. A broadband correlation summary cannot tell the first pair apart, and a summary that discards frequency cannot say which band carries the second difference; the spectral landscape should separate all four. We report empirical Type~I error from $R = 200$ Monte Carlo replications.

\subsection{Simulation}
\label{subsec:sim-design}

\begin{figure}[H]
\centering
\includegraphics[width=.9\linewidth]{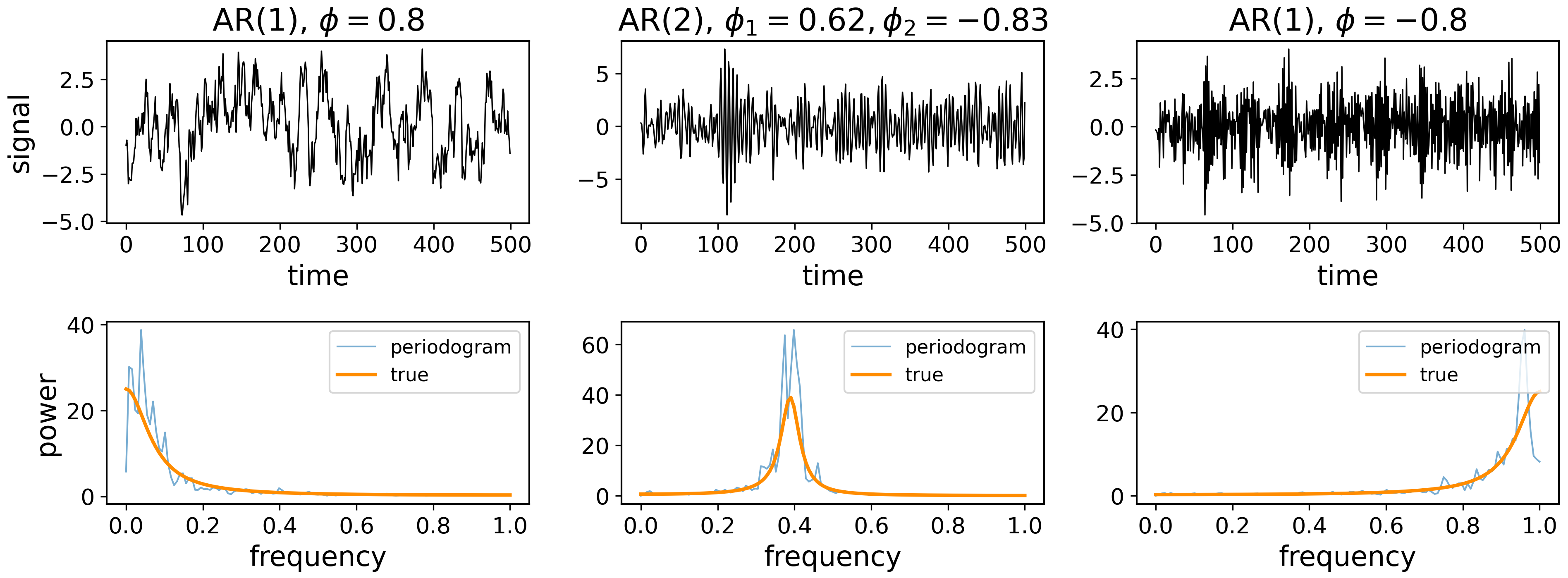}
\caption{Top: three realizations of autoregressive processes used to build dependence at targeted frequency bands. Bottom: corresponding theoretical spectra (orange) and periodogram estimates (blue). Left: AR(1) with $\rho = 0.8$ (low frequency). Center: AR(2) with $\phi_1 = 0.62, \phi_2 = -0.83$ ($\beta$ band). Right: AR(1) with $\rho = -0.8$ (high frequency).}
\label{fig:AR_processes_and_spectra}
\end{figure}

To capture frequency-specific dependence we generate multivariate signals as mixtures of latent autoregressive (AR) processes whose spectra concentrate in target frequency bands, following \citet{EVOLUTIONARY_SSM} and \citet{MIXTURES_AR2}. An AR(1) process with positive coefficient $\rho$ close to $1$ has spectral mass concentrated near zero (low-frequency band); an AR(1) with $\rho$ close to $-1$ has mass concentrated near the Nyquist frequency (high-frequency band); an AR(2) with complex-conjugate roots concentrates mass around a chosen peak frequency (here the $\beta$ band). Figure~\ref{fig:AR_processes_and_spectra} shows three realizations and their spectra.

\begin{figure}[H]
\centering
\includegraphics[width=\linewidth]{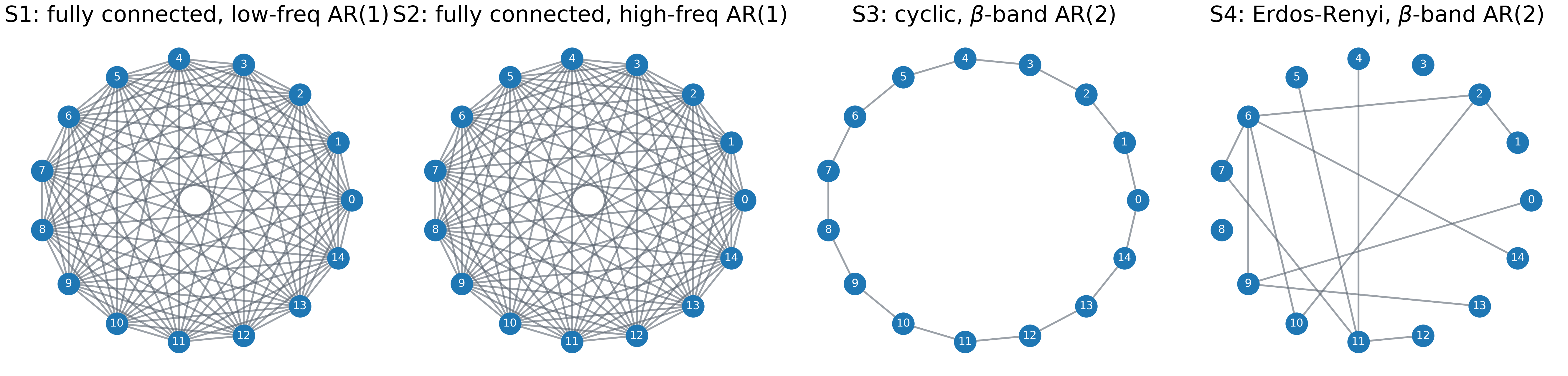}
\caption{The four simulation settings, left to right. S1 and S2 are fully connected, with shared AR(1) dependence concentrated in the low- and high-frequency bands respectively; S3 (cyclic graph) and S4 (Erd\H{o}s--R\'enyi graph) carry shared AR(2) dependence in the $\beta$ band. Nodes are the $P$ channels and edges mark the pairs coupled through a shared latent process.}
\label{fig:simulation_settings}
\end{figure}

We consider four dependence settings, illustrated in Figure~\ref{fig:simulation_settings}:
\begin{enumerate}
\item[\textbf{S1.}] \textbf{Fully connected, low-frequency band:} $X_p(t) = Z_L(t) + \epsilon_p(t)$ for all $p$, where $Z_L$ is a shared AR(1) latent process with $\rho = 0.8$.
\item[\textbf{S2.}] \textbf{Fully connected, high-frequency band:} $X_p(t) = Z_H(t) + \epsilon_p(t)$ for all $p$, where $Z_H$ is a shared AR(1) latent process with $\rho = -0.8$.
\item[\textbf{S3.}] \textbf{Cyclic graph, $\beta$ band:} pairwise dependence following a $P$-cycle topology, induced by shared AR(2) latent processes concentrated in $[12, 27]$~Hz.
\item[\textbf{S4.}] \textbf{Erd\H{o}s--R\'enyi graph, $\beta$ band:} pairwise dependence following a random graph with edge probability $p_E = 2/(P-1)$, matching the expected edge count and mean degree of S3, with the same AR(2) latent structure.
\end{enumerate}

Unless stated otherwise, each setting uses $P = 15$ channels and $T = 1000$ time points, group size $n = 20$, and sampling frequency $f_s = 100$~Hz. Coherence is estimated by a smoothed periodogram, averaging each frequency over a backward window of $m = 5$ Fourier bins, and persistent homology is computed in dimensions $k = 0$ and $k = 1$.

\begin{figure}[H]
\centering
\includegraphics[width=.8\linewidth]{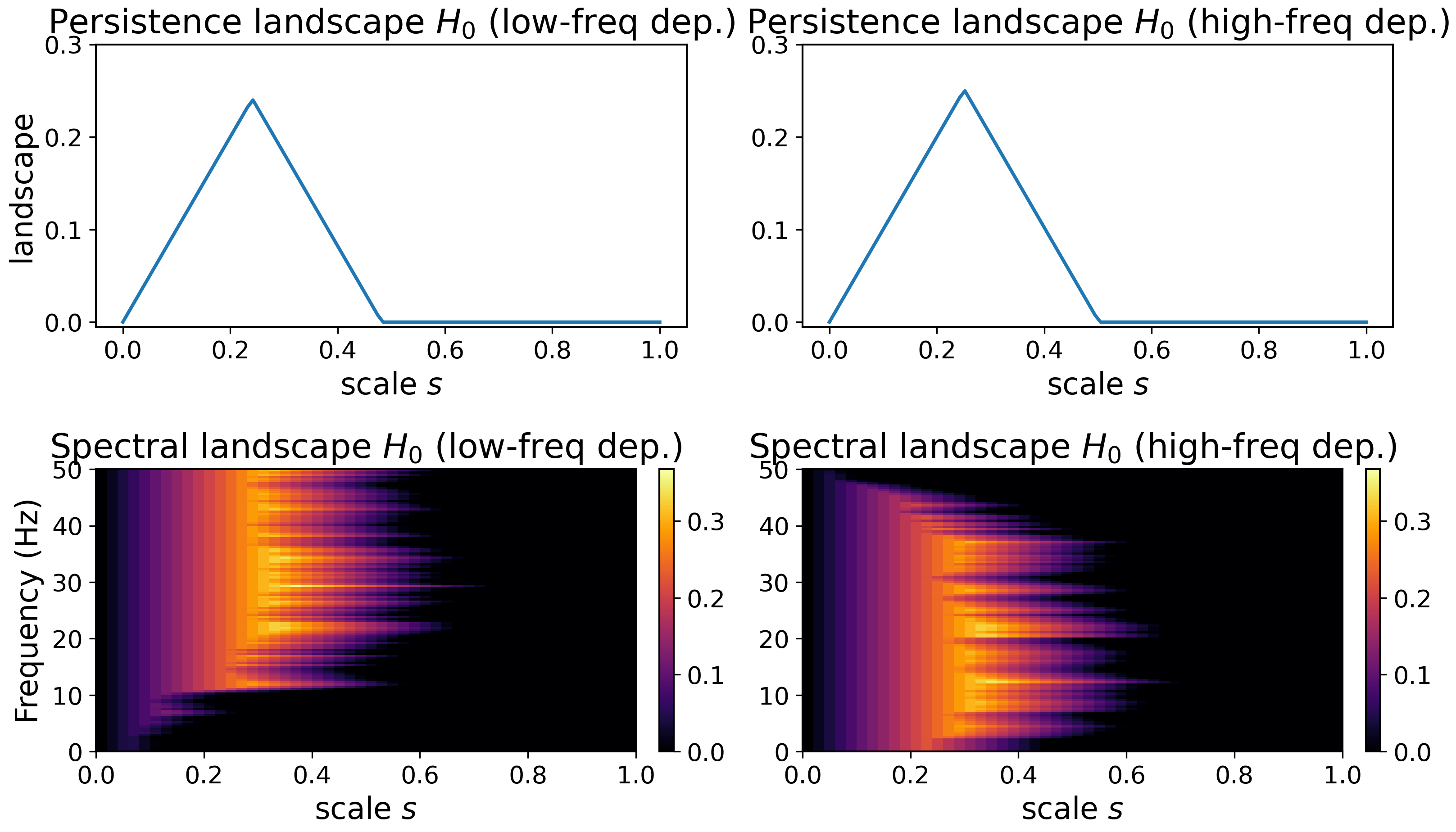}
\caption{Motivating example. Top: traditional persistence landscapes (correlation-based) for two $5$-channel systems differing only in the frequency band at which dependence is concentrated, low (left) and high (right). The two landscapes are nearly identical and cannot distinguish the systems. Bottom: the corresponding spectral landscapes (coherence-based). The spectral information is preserved and the two systems are clearly distinguishable.}
\label{fig:SL_PL}
\end{figure}

The illustrative comparison in Figure~\ref{fig:SL_PL} uses two $5$-channel time series of length $T = 1000$: one with pairwise dependence concentrated in the low-frequency band (mixture of AR(1) with $\rho = 0.8$) and one with the same topology of dependence but concentrated in the high-frequency band (AR(1) with $\rho = -0.8$). Both series produce nearly identical Pearson correlation matrices and therefore nearly identical persistence landscapes (top row), but their spectral landscapes (bottom row) differ sharply because the dependence lives in different parts of the spectrum.

\subsection{Descriptive comparison of spectral landscapes across settings}
\label{subsec:sim-descriptive}

\begin{figure}[H]
\centering
\includegraphics[width=.85\linewidth]{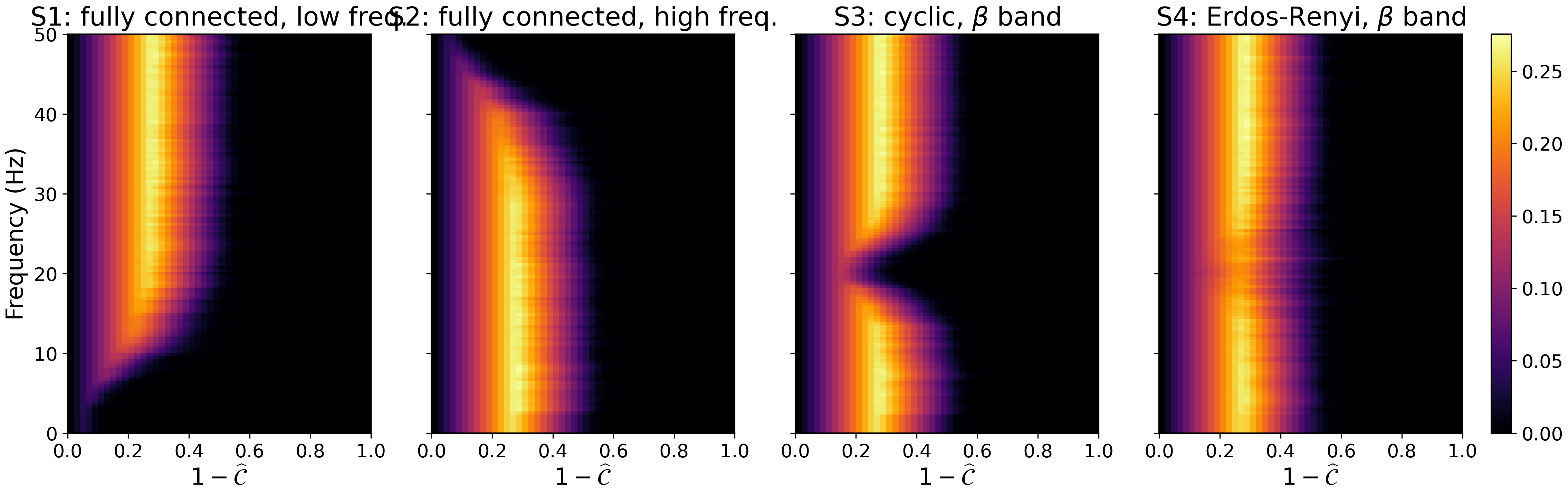}
\caption{Group-average spectral landscapes (left to right: S1--S4) for the 0-dimensional homology group. Horizontal axis: $1 - \widehat{\Cohi}$. Vertical axis: frequency in Hz. Color: landscape height.}
\label{fig:spectral_landscapes_0D}
\end{figure}

\begin{figure}[H]
\centering
\includegraphics[width=.85\linewidth]{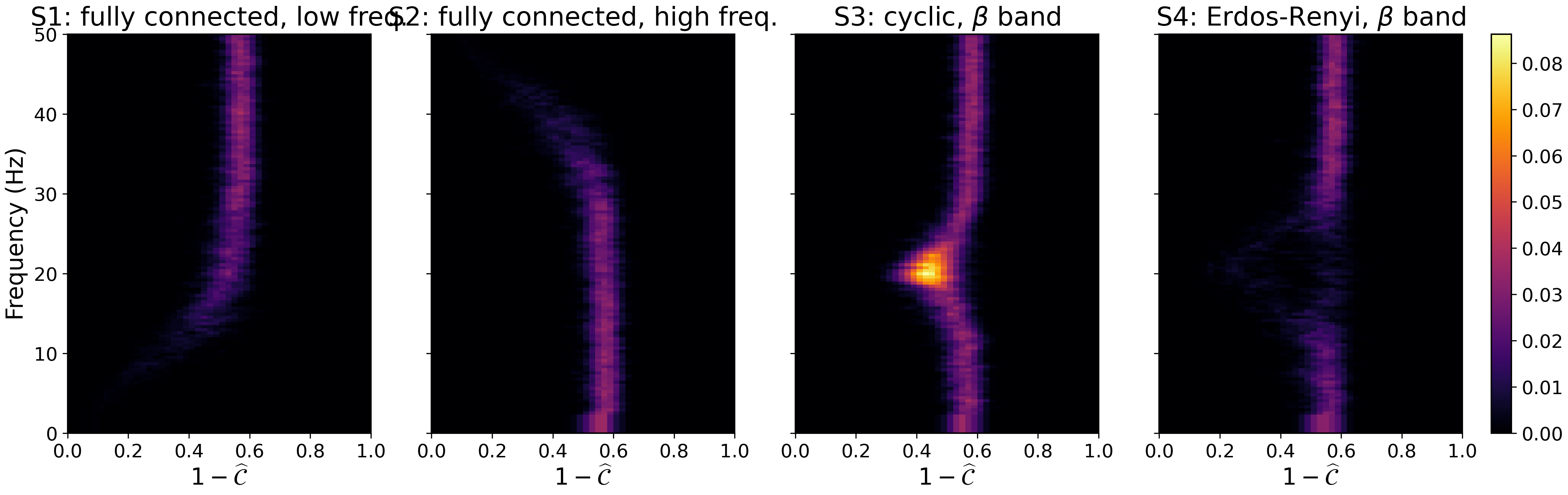}
\caption{Group-average spectral landscapes (left to right: S1--S4) for the 1-dimensional homology group.}
\label{fig:spectral_landscapes_1D}
\end{figure}

We first confirm by eye that the group-average spectral landscape reflects the structure built into each setting. Figures~\ref{fig:spectral_landscapes_0D} and~\ref{fig:spectral_landscapes_1D} show these averages for $k = 0$ (connected components) and $k = 1$ (cycles). The S1 and S2 surfaces separate at the low- and high-frequency tails of the $k = 0$ landscape and coincide through the middle, exactly as expected when full connectivity is moved from one band to another. The S3 and S4 surfaces agree almost everywhere and part only inside the $\beta$ band of the $k = 1$ landscape, the signature of a cyclic-versus-random contrast that is confined to that band.

We then test every pair of settings within each of three frequency bands (low $0$--$12$, middle $12$--$27$, and high $27$--$50$~Hz), a family of $B = 18$ tests per homology dimension ($\binom{4}{2} = 6$ pairs across three bands). Table~\ref{tab:pairwise_testing} reports raw $p$-values from a single realisation of $n = 20$ subjects per group; entries that survive Bonferroni correction at $\alpha = 0.05$ (raw $p < 0.05/18 \approx 0.0028$) are in bold. The two contrasts that isolate a single design axis are listed first: S1 vs S2 (same topology, different band) and S3 vs S4 (same band, different topology).

\begin{table}[H]
\caption{Pairwise testing across the four settings, by frequency band and homology dimension. Entries are raw $p$-values from a single simulation realisation with $n = 20$ subjects per scenario and $B_{\mathrm{MC}} = 10\,000$ Monte Carlo draws from the chi-square mixture null. Each pair is tested in each of the three bands for both $H_0$ (clusters) and $H_1$ (cycles). Bonferroni threshold at $\alpha = 0.05$ across $B = 18$ tests is $p_{\mathrm{Bonf}} \approx 0.0028$; entries below this threshold are in bold. The first two rows isolate a single design axis (frequency for S1 vs S2, topology for S3 vs S4).}
\label{tab:pairwise_testing}
\vspace{2mm}
\begin{center}
\small
\begin{tabular}{l|cc|cc|cc}
\hline
& \multicolumn{2}{c|}{Low: 0--12 Hz} & \multicolumn{2}{c|}{Middle: 12--27 Hz} & \multicolumn{2}{c}{High: 27--50 Hz} \\
Pair & $H_0$ & $H_1$ & $H_0$ & $H_1$ & $H_0$ & $H_1$ \\
\hline
S1 vs S2 & \textbf{0.000} & \textbf{0.000} & \textbf{0.000} & \textbf{0.000} & \textbf{0.000} & \textbf{0.000} \\
S3 vs S4 & 0.488 & 0.010 & \textbf{0.000} & \textbf{0.000} & 0.432 & 0.007 \\
\hline
S1 vs S3 & \textbf{0.000} & \textbf{0.000} & \textbf{0.000} & \textbf{0.000} & 0.110 & \textbf{0.002} \\
S1 vs S4 & \textbf{0.000} & \textbf{0.000} & \textbf{0.000} & \textbf{0.000} & 0.184 & 0.028 \\
S2 vs S3 & 0.032 & 0.006 & \textbf{0.000} & \textbf{0.000} & \textbf{0.000} & \textbf{0.000} \\
S2 vs S4 & 0.026 & \textbf{0.000} & 0.004 & \textbf{0.000} & \textbf{0.000} & \textbf{0.000} \\
\hline
\end{tabular}
\end{center}
\end{table}

The pattern of rejections tracks the design. S1 and S2 carry the same topology in different bands, and they separate in every band and dimension, with raw $p$-values below the Bonferroni threshold throughout: moving full connectivity from the low to the high band changes the coherence across the whole spectrum, not inside a single band. S3 and S4 share the $\beta$ band but differ in graph topology, and they separate cleanly only in the middle band that contains $\beta$ (bold for both $H_0$ and $H_1$), while the low- and high-band contrasts stay above the corrected threshold. A difference that is localised in frequency is detected in the band where it lives and nowhere else, and a difference that is spread across frequency is detected everywhere. The remaining mixed contrasts, which differ in both band and topology at once, separate in most cells as expected. We turn next to the calibration of the test under the null.

\subsection{Type~I error calibration}
\label{subsec:type1}

We assess the empirical Type~I error of the test in~\eqref{eq:teststat} by drawing two groups of $n = 20$ subjects from the same process under each of the four settings, so that the null $H_0: \mu_1 = \mu_2$ holds by construction. For each of $R = 200$ replications, each of the three simulation bands (low, middle, high), and each homology dimension $k \in \{0, 1\}$, we form the statistic $T^{\Omega_\ell}_{N_1, N_2}$, draw $5\,000$ samples from the approximate null $\sum_{d=1}^{D} \widehat{\lambda}_d N_d^2$ to set the critical value $\widehat{\tau}_\ell$ at $\alpha = 0.05$, and record a rejection when $T^{\Omega_\ell}_{N_1, N_2} > \widehat{\tau}_\ell$. The rejection rate over the $R$ replications estimates the Type~I error; under correct calibration it should sit near $0.05$.

\begin{table}[H]
\centering
\caption{Empirical Type~I error rates at nominal level $\alpha = 0.05$, for two groups of $n = 20$ subjects drawn from the same process, across the four settings, the three simulation bands, and the two homology dimensions ($R = 200$ Monte Carlo replications). The Monte Carlo standard error is $\sqrt{0.05 \cdot 0.95 / 200} \approx 0.015$.}
\label{tab:type1}
\vspace{2mm}
\begin{center}
\small
\begin{tabular}{ll|ccc}
\hline
Setting & Dim. & low (0--12 Hz) & middle (12--27 Hz) & high (27--50 Hz) \\
\hline
\multirow{2}{*}{S1: fully conn., low}            & $H_0$ & 0.065 & 0.045 & 0.010 \\
                                                  & $H_1$ & 0.025 & 0.010 & 0.000 \\
\hline
\multirow{2}{*}{S2: fully conn., high}           & $H_0$ & 0.060 & 0.035 & 0.020 \\
                                                  & $H_1$ & 0.050 & 0.020 & 0.000 \\
\hline
\multirow{2}{*}{S3: cyclic, $\beta$}             & $H_0$ & 0.075 & 0.040 & 0.025 \\
                                                  & $H_1$ & 0.020 & 0.015 & 0.005 \\
\hline
\multirow{2}{*}{S4: Erd\H{o}s--R\'enyi, $\beta$} & $H_0$ & 0.065 & 0.035 & 0.030 \\
                                                  & $H_1$ & 0.030 & 0.015 & 0.005 \\
\hline
\end{tabular}
\end{center}
\end{table}

Across the $24$ cells, $19$ fall at or below the nominal $0.05$ and the largest is $0.075$ (S3, $k = 0$, low band), about $1.7$ standard errors above nominal. The few rates above $0.05$ all sit in the low-frequency $H_0$ block, where the coherence estimator carries its largest variance. Elsewhere the test is mildly conservative, more so for $H_1$ than for $H_0$ and for the narrow high-frequency band, which is consistent with the smaller effective dimension of the pooled covariance in those cells. The test holds its level across settings, bands, and dimensions.

\section{Application}
\label{sec:app}

Attention deficit hyperactivity disorder (ADHD) is one of the most common neurodevelopmental conditions of childhood, and its electrophysiological correlates have been studied for decades through band-specific measures of EEG power and pairwise coupling \citep{ADAMOU_ADHD_EEG_REVIEW, MICHELINI_ADHD_OSCILLATIONS, SU_ADHD_EEG_REVIEW}. Such measures record how strongly pairs of channels co-vary within a rhythm, but they say little about how that synchrony is \emph{organised} across the montage: whether the coherence network at a given rhythm clusters into tight communities or threads into closed feedback loops, and whether that organisation differs between groups.

\subsection{Data and preprocessing}
\label{subsec:data}

\begin{figure}[H]
\centering
\begin{minipage}[t]{0.28\columnwidth}
\includegraphics[width=\linewidth]{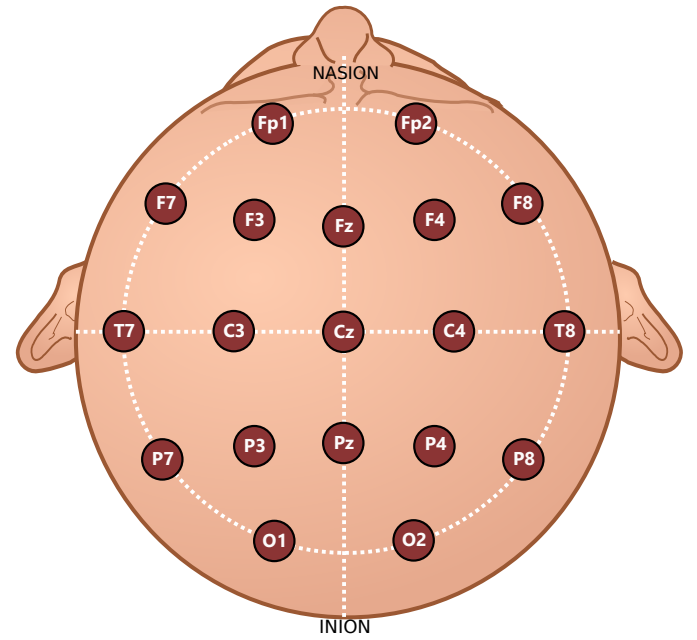}
\subcaption{10--20 scalp layout.}
\label{fig:scalp_eeg}
\end{minipage}\hfill
\begin{minipage}[t]{0.71\columnwidth}
\includegraphics[width=\linewidth]{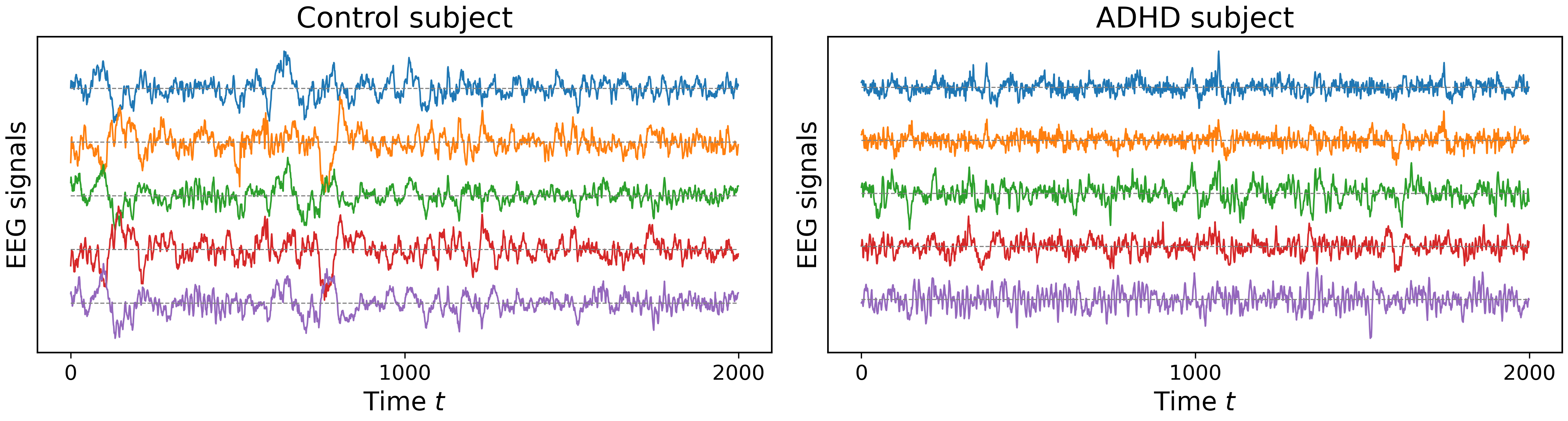}
\subcaption{Example preprocessed signals (five electrodes), control (left) and ADHD (right).}
\label{fig:EEG_Signals}
\end{minipage}
\caption{EEG data set: electrode layout and example recordings.}
\label{fig:eeg-data}
\end{figure}

We use the publicly available ADHD EEG data set of \citet{ADHD_DATA}, comprising EEG recordings from $61$ children diagnosed with ADHD and $60$ healthy controls, aged $7$--$12$ years. The ADHD group had received methylphenidate (Ritalin) for up to six months prior to recording. None of the controls had a history of psychiatric disorders, epilepsy, or high-risk behaviors. EEG was acquired with $19$ scalp electrodes in the international 10--20 layout \citep{BRAIN_EEG} at a sampling frequency of $f_s = 128$~Hz (Figure~\ref{fig:scalp_eeg}). Recordings were collected during a visual attention task in which subjects were shown sets of characters and asked to count them; recording duration varied by response time. After preprocessing with the PREP pipeline \citep{EEG_PREP, ADHD_DATA} (electrical-line removal, ocular and muscular artifact removal, bad-channel detection and repair, filtering, and re-referencing), $N_1 = 53$ control and $N_2 = 51$ ADHD subjects remained.

The two groups have overlapping age ranges (7--12 years), and recording duration varied by subject due to the self-paced task design. The data release \citep{ADHD_DATA} reports a mixed cohort of boys and girls but does not provide a per-subject sex label, so the groups cannot be matched on sex. The ADHD children had taken methylphenidate for up to six months prior to recording, though their medication state during the recording session is not specified.

\subsection{Estimation pipeline}
\label{subsec:estimation}

For each subject $i$, we estimate the spectral density matrix $\widehat{S}^{(i)}(\omega)$ by smoothed periodogram on the full preprocessed recording, averaging each frequency over a backward window of $m = 7$ Fourier bins. This bandwidth is a standard default for clinical EEG at $f_s = 128$~Hz. The coherence $\widehat{\Cohi}^{(i)}_{pq}(\omega)$ is then computed from~\eqref{eq:coherence} on the full Fourier grid up to the Nyquist frequency $f_s / 2 = 64$~Hz ($129$ frequency points).

Coherence is integrated within each of the five canonical EEG bands:
$\delta = [0.5, 4]$~Hz, $\theta = [4, 8]$~Hz, $\alpha = [8, 12]$~Hz, $\beta = [12, 30]$~Hz, $\gamma = [30, 50]$~Hz. The choice of bands follows long-standing clinical convention \citep{COHERENCE_HIST, ABHANG_2016}; the spectral landscape itself is computed on the full Fourier grid and the band structure enters only at the testing stage via the choice of $\Omega_\ell$ in~\eqref{eq:H0}. For each subject $i$, frequency $\omega$ on the Fourier grid, and homology dimension $k \in \{0, 1\}$, we compute the Vietoris-Rips filtration of the $P \times P$ distance matrix $1 - \widehat{\Cohi}^{(i)}(\omega)$ on $P = 19$ channels, extract the persistence diagram $\dgm_k^{(i)}(\omega)$ with the Ripser algorithm \citep{BAUER_RIPSER}, and assemble the spectral landscape $\Lspl_k^{(i)}(s, \omega)$ on a grid of $50$ filtration scales and the $129$ Fourier frequencies.

\subsection{Population spectral landscapes}
\label{subsec:pop-SL}

\begin{figure}[H]
\centering
\begin{minipage}{.48\textwidth}
\centering
\includegraphics[width=\linewidth]{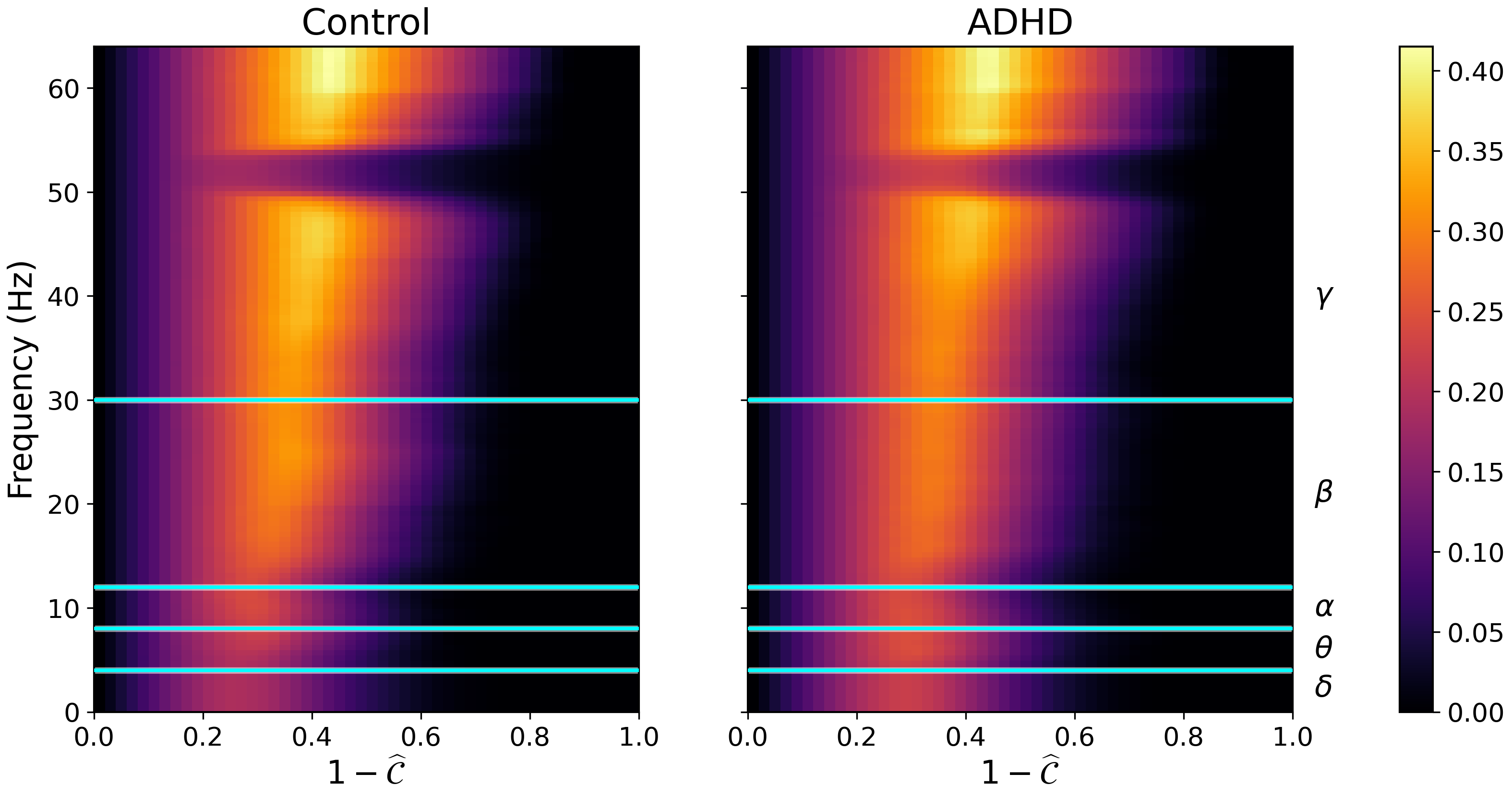}
\subcaption{$H_0$: connected components.}
\label{fig:persistance_surfaces_0_Homology}
\end{minipage}\hfill
\begin{minipage}{.48\textwidth}
\centering
\includegraphics[width=\linewidth]{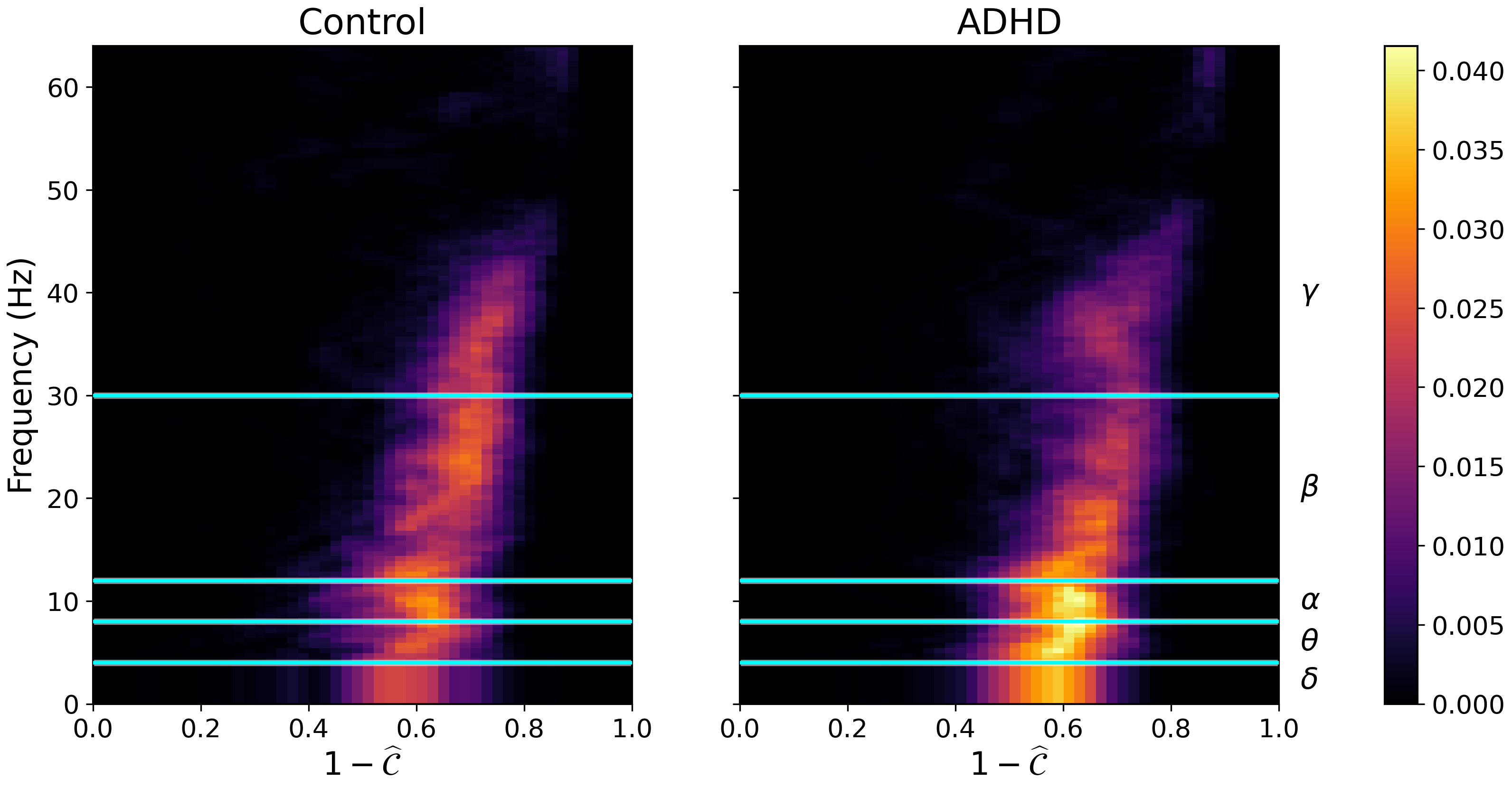}
\subcaption{$H_1$: cycles.}
\label{fig:persistance_surfaces_1_Homology}
\end{minipage}
\caption{Population-mean spectral landscapes for control (left in each panel) and ADHD (right): (a)~$k = 0$ (components), (b)~$k = 1$ (cycles). Cyan lines mark the canonical EEG band edges; the horizontal axis is $1 - \widehat{\Cohi}$, the vertical axis frequency.}
\label{fig:pop-SL}
\end{figure}

Figure~\ref{fig:pop-SL} shows the sample-mean spectral landscapes for control ($\widehat{\mu}_1$) and ADHD ($\widehat{\mu}_2$), for $k = 0$ and $k = 1$. The vertical axis is frequency and the horizontal axis is the filtration scale $1 - \widehat{\Cohi}$, so the left edge is high coherence and the right edge low coherence; the surface height is the persistence value, larger where a feature survives over a wider range of scales. The $H_0$ surface (components) is a broad, tall ridge at intermediate coherence ($1 - \widehat{\Cohi} \approx 0.3$--$0.5$), strongest through the $\beta$ and $\gamma$ bands. The $H_1$ surface (cycles) is narrower and lower and runs diagonally, from intermediate scales in $\delta$--$\theta$ to larger scales through $\beta$ and into mid-$\gamma$, fading at the top of $\gamma$, so synchronisation loops close at progressively weaker coherence as frequency rises. Control and ADHD differ most in the $\beta$--$\gamma$ reach of the $H_0$ ridge and in the $\theta$ and $\gamma$ portions of the $H_1$ ridge. These are averages over more than fifty subjects per group, so the differences are not single-recording artefacts; they motivate the principal-component view next and the formal test of Section~\ref{subsec:freq-test}.

\subsection{Functional principal-component decomposition}
\label{subsec:fpca}

Before the formal test, we look at the structure of the data through a functional principal-component analysis (FPCA) of the control-group spectral landscapes. Following \citet[Ch.~3]{horvath2012inference}, we compute the empirical covariance kernel $\widehat{\Gamma}_1$ on the centred control landscapes, extract its leading eigenfunctions $\{\widehat{\phi}_d\}_{d \ge 1}$ by Mercer's theorem \citep{STEINWART_KERNELS}, and project every subject, control and ADHD alike, onto the resulting components. This shows where between-group variability concentrates in the $(s,\omega)$ plane, and it checks whether the groups separate at all in a low-dimensional projection before we commit to a fully functional test.

\begin{figure}[H]
\centering
\includegraphics[width=\linewidth]{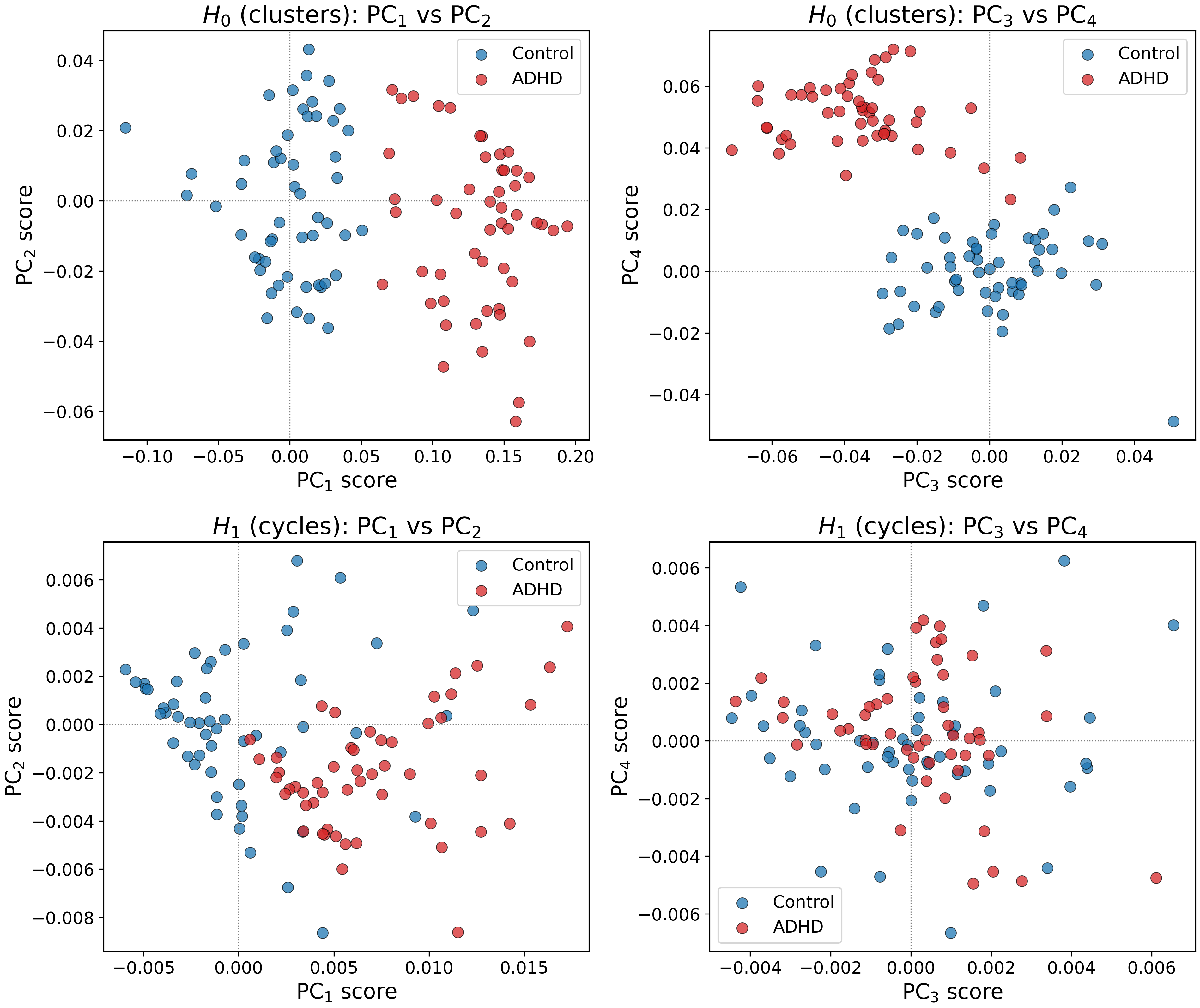}
\caption{Subject scores on the leading control-group principal components. Rows: $k = 0$ (top), $k = 1$ (bottom); columns: $\text{PC}_1$--$\text{PC}_2$ (left), $\text{PC}_3$--$\text{PC}_4$ (right). Control blue, ADHD red. The control group is centred for the fit; ADHD subjects are projected raw, so level differences remain in the scores.}
\label{fig:fpca-scatter}
\end{figure}

Figure~\ref{fig:fpca-scatter} projects all $N_1 + N_2 = 104$ subjects onto the first four control-group components, for $k = 0$ and $k = 1$, in a single $2 \times 2$ grid (rows: topology; columns: PC pair). For $k = 0$ (top row) the groups separate along $\text{PC}_1$: controls (blue) cluster near $\text{PC}_1 \approx 0$ while ADHD subjects (red) shift to $\text{PC}_1 \approx 0.13$, and the $\text{PC}_3$-versus-$\text{PC}_4$ panel adds a second axis, with ADHD scores on $\text{PC}_4$ sitting above zero. For $k = 1$ (bottom row) the separation is weaker: the clouds overlap more along $\text{PC}_1$ and there is no comparable axis in $\text{PC}_3$ versus $\text{PC}_4$. This separation is recovered without ever using the diagnostic label, which is what we would expect if the two groups genuinely differ in the topology of their coherence networks. The gap between the rows, sharper in the cluster summary $H_0$ and subtler in the cycle summary $H_1$, recurs through the rest of the analysis and appears again in the test $p$-values below.

\begin{figure}[H]
\centering
\includegraphics[width=.95\linewidth]{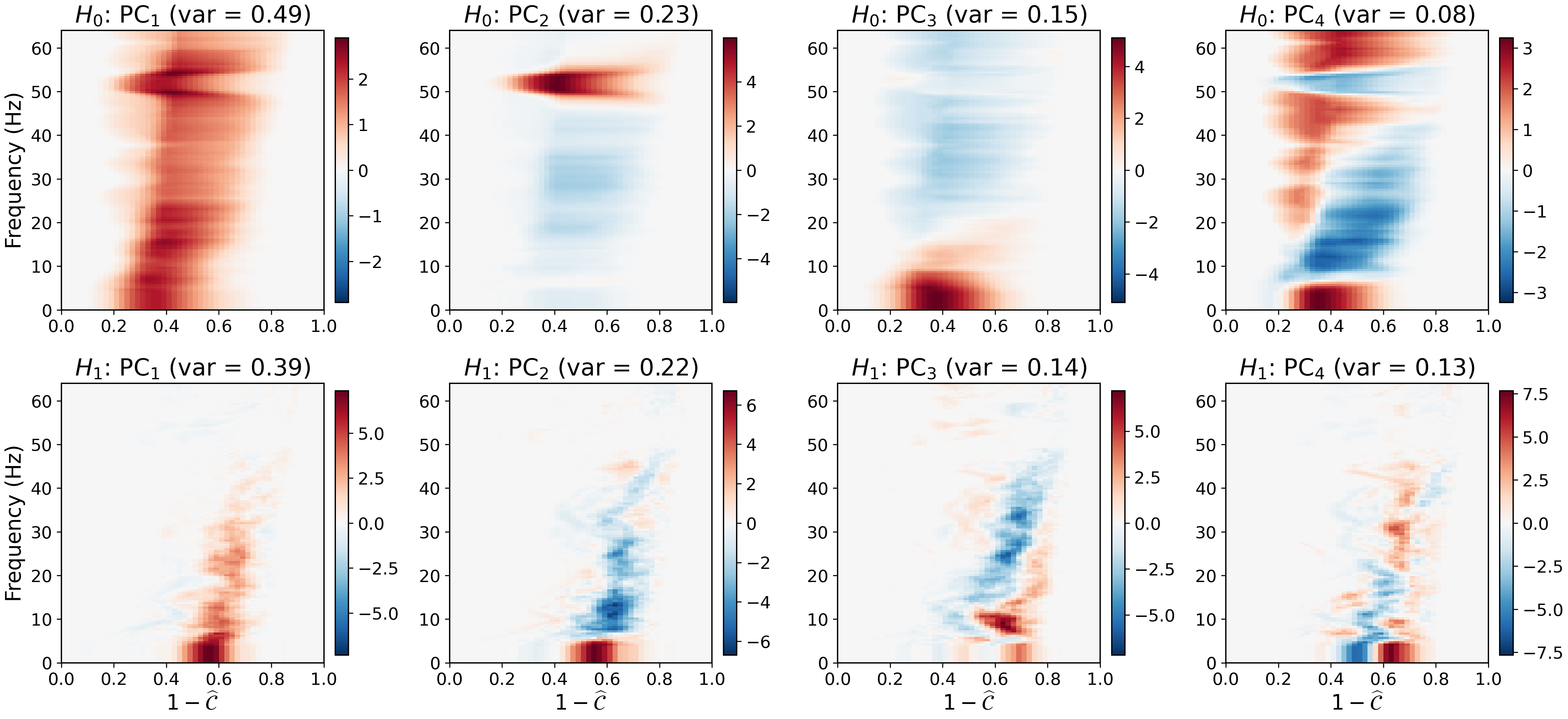}
\caption{Leading eigenfunctions $\widehat{\phi}_1, \dots, \widehat{\phi}_4$ of the control-group $k = 0$ covariance (frequency vertical, scale $s$ horizontal); red and blue mark positive and negative weight. Explained-variance ratios above each panel.}
\label{fig:fpca-h0-eigfuns}
\end{figure}

Figure~\ref{fig:fpca-h0-eigfuns} shows the four leading $k = 0$ eigenfunctions: $\text{PC}_1$ is a broad, mostly positive mode peaking near $50$~Hz; $\text{PC}_2$ contrasts $\gamma$-band against $\beta$-band weight; $\text{PC}_3$ contrasts low-frequency ($\delta$/$\theta$) against the rest of the spectrum; and $\text{PC}_4$ captures finer multi-band variation.

While the FPCA visualization is useful as exploratory evidence, we follow \citet{Functional_Testing_1} and \citet{Functional_Testing_2} in preferring a fully functional test that does not rely on a finite-dimensional projection, since power can be lost when the leading principal components fail to align with the difference $\mu_1 - \mu_2$. The formal test below uses the full functional structure of the spectral landscape.

\subsection{Frequency-specific testing with multiple-testing correction}
\label{subsec:freq-test}

We first ask whether the two groups differ at all in the topology of their coherence networks, before asking where. Pooling the whole analysis range into a single band $\Omega_{\mathrm{global}} = [0.5, 50]$~Hz and applying the test of Theorem~\ref{thm:test} once per homology dimension, with the same statistic and the same $50\,000$ Monte Carlo null draws used below, the cycle topology shows a significant global difference: for $k = 1$, the observed statistic $T_{\mathrm{obs}} = 1.17$ exceeds its $95\%$ critical value $q_{0.95} = 0.99$ (raw $p = 0.019$), so we reject $H_0: \mu_1 = \mu_2$ over the whole band at the $95\%$ level. For the connected-component topology the global test is borderline and does not reach significance ($k = 0$: $T_{\mathrm{obs}} = 28.44$, $q_{0.95} = 29.09$, raw $p = 0.053$). The one claim we make at the $95\%$ level is therefore this global difference in cycle topology; the band-by-band analysis that follows asks where in the spectrum it sits and is exploratory.

To localise the effect, we apply the same test to each of the five canonical EEG bands and each of the two homology dimensions, a family of $B = 10$ tests, and report raw $p$-values alongside Bonferroni-corrected $p$-values (multiplying by $B = 10$, capping at $1$) and Benjamini--Hochberg-adjusted $p$-values at FDR level $q = 0.05$. The reference null distribution is the same as for the global test, approximated by $50\,000$ Monte Carlo draws from $\sum_{d=1}^{D} \widehat{\lambda}_d N_d^2$ with $D = N_1 + N_2 - 2 = 102$. Because these ten tests are a follow-up to the global rejection rather than independent confirmatory analyses, we read them as exploratory localisation.

Table~\ref{tab:adhd-pvals} collects the global test and the raw and adjusted $p$-values for all ten band-by-dimension tests.

\begin{table}[H]
\centering
\caption{Frequency-specific test of $H_0: \mu_1 = \mu_2$, control vs.\ ADHD ($N_1 = 53$, $N_2 = 51$). The \emph{Global} row pools the whole analysis band $[0.5, 50]$~Hz into a single test per dimension, so no multiplicity correction applies. The per-band rows form the localisation family of $B = 10$ tests, with Bonferroni (raw $p \times 10$, capped at $1$) and Benjamini--Hochberg adjusted values. Raw $p$-values are from Theorem~\ref{thm:test}; entries with raw $p < 0.05$ are in bold.}
\label{tab:adhd-pvals}
\vspace{2mm}
\begin{center}
\begin{tabular}{l|ccc|ccc}
\hline
 & \multicolumn{3}{c|}{$k = 0$ (clusters)} & \multicolumn{3}{c}{$k = 1$ (cycles)} \\
Band & raw & Bonf & BH & raw & Bonf & BH \\
\hline
Global $[0.5, 50]$~Hz & 0.053 & $-$ & $-$ & \textbf{0.019} & $-$ & $-$ \\
\hline
$\delta$  & 0.092 & 0.917 & 0.153 & 0.131 & 1.000 & 0.187 \\
$\theta$  & \textbf{0.044} & 0.445 & 0.111 & \textbf{0.010} & 0.103 & 0.060 \\
$\alpha$  & 0.413 & 1.000 & 0.413 & 0.075 & 0.755 & 0.151 \\
$\beta$   & 0.262 & 1.000 & 0.291 & 0.213 & 1.000 & 0.266 \\
$\gamma$  & \textbf{0.025} & 0.245 & 0.082 & \textbf{0.012} & 0.119 & 0.060 \\
\hline
\end{tabular}
\end{center}
\end{table}

At the unadjusted level, four of the ten band-dimension combinations are significant at $\alpha = 0.05$: $\theta$ band for $k = 0$ ($p = 0.044$), $\gamma$ band for $k = 0$ ($p = 0.025$), $\theta$ band for $k = 1$ ($p = 0.010$), and $\gamma$ band for $k = 1$ ($p = 0.012$). None of these survive Bonferroni correction at the family-wise level $\alpha = 0.05$; the smallest Bonferroni-adjusted $p$-value is $0.103$ ($\theta$ band, $k = 1$). Under Benjamini--Hochberg control of the false discovery rate at $q = 0.05$, no test reaches significance, but the $\theta$ and $\gamma$ bands for $k = 1$ are tied for the smallest adjusted $p$-value ($p_{\mathrm{BH}} = 0.060$), followed by $\gamma$ band for $k = 0$ ($p_{\mathrm{BH}} = 0.082$) and $\theta$ band for $k = 0$ ($p_{\mathrm{BH}} = 0.111$).

The global $k = 1$ test has already established a difference at the $95\%$ level; the per-band picture says that difference is carried mainly by the $\theta$ and $\gamma$ bands, and the same two bands stand out in the $k = 0$ summary. We read the individual bands as localisation of the global effect rather than as ten separately confirmed tests, since none survives strict family-wise correction at this sample size. A larger study (a few hundred subjects per group) would let the band-level effects, and the borderline global $k = 0$ result, be tested in their own right.

\subsection{Biological interpretation}
\label{subsec:bio}

The topological summaries used here have direct neurophysiological readings. The 0-dimensional homology $H_0$ tracks connected components of the coherence-thresholded network: at a given filtration scale $s$, the number of connected components corresponds to the number of communities of channels that share strong synchrony at frequency $\omega$ at that scale. Changes in the $H_0$ spectral landscape between groups therefore reflect changes in the \emph{clustering} of channels into functionally synchronized communities. The 1-dimensional homology $H_1$ tracks one-dimensional cycles (loops) in the coherence graph: a cycle corresponds to a set of channels arranged so that synchrony forms a closed feedback path with no internal cross-connection (otherwise the cycle would be filled in by a higher simplex). Changes in the $H_1$ spectral landscape between groups therefore reflect changes in the presence and persistence of such feedback or re-entrant loops at frequency $\omega$.

The $\gamma$-band differences we see in both $H_0$ and $H_1$ are consistent with the established literature linking $\gamma$ oscillations to attentional control, working memory, and information binding \citep{ABHANG_2016}, and with the literature linking $\gamma$ abnormalities specifically to ADHD. \citet{MICHELINI_ADHD_OSCILLATIONS} review event-related oscillatory differences in ADHD across bands, and \citet{TOMBOR_GAMMA_ADHD} document atypical gamma trajectories in adult ADHD at rest. Our results add that, beyond the well-documented changes in $\gamma$-band power and pairwise coupling, the higher-order topology of $\gamma$-band coherence networks is altered in ADHD, both in the clustering pattern of channels ($H_0$) and in the presence of synchronised feedback loops ($H_1$). The $\theta$-band differences in both $H_0$ and $H_1$ are consistent with the role of $\theta$ rhythms in attention regulation and with the long-standing observation that ADHD is associated with elevated $\theta$ activity at rest \citep{MICHELINI_ADHD_OSCILLATIONS}. The global difference in cycle topology is established at the $95\%$ level; the specific band attributions, by contrast, do not survive strict family-wise control at this sample size and should be read as localisation to be confirmed in a larger replication study.

\section{Conclusion}
\label{sec:conc}

We introduced the spectral landscape, a two-dimensional topological summary that indexes the persistence landscape of \citet{PL_FIRST} by Fourier frequency, together with a functional two-sample test for differences between population means over a chosen frequency band. The construction is Lipschitz-stable in the underlying coherence matrix (Theorem~\ref{thm:stability-pointwise}) and the test has the asymptotic null distribution given in Theorem~\ref{thm:test}, with consistency following from Proposition~\ref{prop:limits}.

On the ADHD EEG data of \citet{ADHD_DATA}, a global test over the full analysis band rejects equality of the two groups' cycle topology at the $95\%$ level ($p = 0.019$), establishing that ADHD and control children differ in the topology of their brain coherence networks. A band-by-band follow-up localises this difference to the $\gamma$ and $\theta$ bands, in both connected-component and cycle summaries, although no individual band survives strict family-wise correction over the $B = 10$ tests at the current sample size. The localisation is consistent with the established neuroscience literature on $\gamma$- and $\theta$-band involvement in attention and in ADHD.


\bigskip
\begin{center}
{\large\bf SUPPLEMENTAL MATERIALS}
\end{center}
\begin{description}
\item[Supplement A. Background on persistence homology:] Self-contained
review of simplicial complexes, Vietoris--Rips filtrations, Betti numbers, barcodes,
persistence diagrams, and persistence landscapes for readers unfamiliar with
TDA.
\item[Supplement B. Additional proofs:] Detailed proofs of the asymptotic
limit theorems (Section~\ref{subsec:asymptotics}), the stability theorem
(Section~\ref{subsec:stability}), and the asymptotic null distribution of the
test (Section~\ref{subsec:test}).
\end{description}





\vspace{-3mm}
\bibliographystyle{apalike}  
\bibliography{references}

\end{document}